\documentclass[%
 reprint,
 amsmath,amssymb,
 aps,
]{revtex4-2}
\usepackage{ragged2e}
\usepackage{lineno}
\usepackage{amssymb,amsmath,amsthm,enumitem}
\usepackage{bigints}
\usepackage{array}
\usepackage[autostyle]{csquotes}
\usepackage[british]{babel}
\usepackage{graphicx}
\usepackage{dcolumn}
\usepackage{bm}
\usepackage[caption=false]{subfig}
\makeatletter
\renewcommand{\fnum@figure}{FIG. \thefigure}
\renewcommand{\fnum@table}{TABLE \thetable}
\makeatother
\usepackage{floatrow}
\usepackage{epstopdf}
\begin{document}
\setlength{\abovedisplayskip}{5pt}
\setlength{\belowdisplayskip}{5pt}
\preprint{APS/123-QED}

\title{Analysis of the Mobility-Limiting Mechanisms of the Two-Dimensional Hole Gas on Hydrogen-Terminated Diamond}

\author{Ricardo Peterson$^1$}\thanks{Author to whom correspondence should be addressed. Electronic mail:rp3@stanford.edu}
\def\andname{,}
\author{Mohamadali Malakoutian$^2$, Xiaoqing Xu$^3$, Caitlin Chapin$^4$, Srabanti Chowdhury$^{1,2}$, Debbie G. Senesky$^4$}
 \affiliation{\Centering{$^1$Department of Electrical Engineering, Stanford University, California 94305, USA\\
 $^2$Department of Electrical and Computer Engineering, University of California, Davis, California 95616, USA\\
 $^3$Stanford Nanofabrication Facility, Stanford University, California 94305, USA\\
 $^4$Department of Aeronautics and Astronautics, Stanford University, California 94305, USA}}

\begin{abstract}
Here we present an analysis of the mobility-limiting mechanisms of a two-dimensional hole gas on hydrogen-terminated diamond surfaces. The scattering rates of surface impurities, surface roughness, non-polar optical phonons, and acoustic phonons are included. Using a Schr{\"o}dinger/Poisson solver, the heavy hole, light hole, and split-off bands are treated separately. To compare the calculations with experimental data, Hall-effect structures were fabricated and measured at temperatures ranging from 25 to 700~K, with hole sheet densities ranging from 2 to 6$\times10^{12}\;\text{cm}^{-2}$ and typical mobilities measured from 60 to 100~cm$^{2}$/(V$\cdot$s) at room temperature. Existing data from literature was also used, which spans sheet densities above 1$\times10^{13}\;\text{cm}^{-2}$. Our analysis indicates that for low sheet densities, surface impurity scattering by charged acceptors and surface roughness are not sufficient to account for the low mobility. Moreover, the experimental data suggests that long-range potential fluctuations exist at the diamond surface, and are particularly enhanced at lower sheet densities. Thus, we propose a second type of surface impurity scattering which is caused by disorder related to the C-H dipoles. 
\end{abstract}

\maketitle


\section{Introduction}

As a semiconductor material, diamond has exceptional figures of merit due to its wide band gap, high breakdown voltage, high thermal conductivity, and high carrier mobility~\cite{Geis_2018}. The combination of wide band gap and high electron and hole mobilities is rare among semiconductor materials, which makes diamond an attractive candidate for high power electronics. However, doping of diamond has been a challenge, owing to its large activation energies for dopants (0.37 eV for $\it{p}$-type)~\cite{Collins_1971}, where one in $10^4$ boron dopants is activated at room temperature. For this reason, hydrogen-terminated diamond (H:Diamond) has been studied as an alternative conduction mechanism. It has been demonstrated that when H:Diamond is exposed to air, atmospheric molecules adsorb onto the surface and induce a two-dimensional hole gas (2DHG), achieving a hole density of $10^{12} - 10^{13}$ cm$^{-2}$ and a hole mobility of $50-150$~cm$^2$/(V$\cdot$s)~\cite{Hirama_2010,verona_2016}. An electrochemical surface transfer doping model is most commonly invoked to explain this surface conduction mechanism \cite{Maier_2000}. In this model, electron transfer occurs from the top of the surface diamond's valence band to lower accessible energy states in the atmospheric adsorbates. This causes an alignment of the Fermi energy, which is near or below the top of valence band at the H:Diamond surface. Thus, a hole gas is induced, and a compensating sheet of negative charge is formed in the first mono-layer of the air-adsorbates \cite{KAWARADA1996,Tsugawa_2001}.

At temperatures exceeding $\sim$60$^{\circ}$C, however, the air-adsorbates begin to thermally desorb from the diamond surface, thus causing the 2DHG to collapse \cite{Riedel_2004, verona_2016}. This is unfortunate since the value of diamond-based electronics stems from its potential to operate robustly at high temperatures. For this reason, surface passivation of H:Diamond has been explored as a solution. It was found by Kawarada \textit{et al}. that Al$_2$O$_3$ passivation stabilizes the hole conduction above $400^{\circ}$C \cite{Kawarada_2014}. Since then, other dielectric layers such as HfO$_2$ have been used to passivate the H:Diamond surface \cite{Liu_2013}. Moreover, transition-metal oxides (TMOs) with high work functions such as WO$_3$, V$_2$O$_5$, and MoO$_3$ have been shown to act as efficient electron acceptors \cite{verona_2016,Russel_2013}, thus inducing much higher 2DHG densities.

Despite the successful efforts to increase the hole sheet density, significant limitations to the hole mobility make it challenging to increase the overall conductivity. Hole mobilities well below 150 cm$^2$/(V$\cdot$s) are usually reported \cite{Rezek_2006,verona_2016,Hirama_2010}, which is significantly less than the bulk mobility values. One cause of this is the Coulomb interactions between the 2DHG and the compensating negative charge (i.e., the ionized surface acceptors). This induces significant scattering, particularly at low-to-intermediate temperatures~\cite{Rezek_2006,YLIetal}. Moreover, this scattering mechanism is exacerbated as the sheet density increases, as evidenced by the reduction in mobility for the TMOs with higher work functions \cite{verona_2016}. Additionally, other surface-related phenomena such as incomplete H-termination has been invoked to explain the temperature-dependent behavior of the hole mobility \cite{Nebel_2002, Garrido_2005}. Much like the charged surface acceptors, such irregularities related to the hydrogenated surface would also induce potential fluctuations along the 2D well and thus act as scattering centers.

Calculation of carrier relaxation times as a means to determine the mobility adds great insight to the conduction-limiting mechanisms in any semiconductor technology. Physical insight is crucial from the standpoint of design solutions, which is much needed for immature technologies with great potential such as diamond. Thus, in this paper, we develop a scattering model for hole gases in H:Diamond, where the heavy hole, light hole, and split off bands are treated separately. We then study the effects of different scattering mechanisms over a wide range of temperatures and carrier concentrations. This includes an analysis of two types of surface impurity scattering. The first being via negatively charged surface acceptors (a consequence of the 2DHG formation), and the second being via disorder related to C-H dipoles (a consequence of surface treatment throughout the fabrication process). The latter mechanism can ultimately explain why hole mobilities remain low even at low sheet densities. The final calculations have been fitted to experimental measurements made on fabricated Hall-effect devices, which agree well over a wide range of temperatures and sheet densities.

\section{Experiment}
Here we discuss the experimentally determined variables of interest from the 2DHG in H:Diamond. The measured properties are the 2DHG mobility, sheet density, and surface roughness parameters.

To determined the electrical properties of the 2DHG, Hall measurements were performed on devices with a van der Pauw geometry. The devices were fabricated on four samples of CVD-grown 250~$\mu$m-thick single-crystal (001) diamond, obtained from Element Six Ltd. The diamond surfaces were treated using a hydrogen-plasma in a microwave CVD reactor, resulting in a hydrogen-terminated surface. The plasma power and pressure were 1.35~kW and 30 torr, respectively. This treatment lasted for 30 min with a surface temperature measured at 910$^{\circ}~\pm~10^{\circ}$C.

During the fabrication process, for every lithography step, the diamond surface was only exposed to the LOL2000 solution for lift-off, as well as standard developers and solvents for cleaning. Given the reasonable Hall-effect results taken after fabrication, none of these chemicals are believed to have compromised the H-terminated surface. The Hall-effect devices were fabricated as follows. 
(i) Ti/Pt/Au (5/20/20 nm) bond pads were patterned and deposited via e-beam evaporation, followed by the standard lift-off technique. To ensure good adhesion and ease of wirebonding, the patterned bond pad regions were oxygen-terminated in a 100W O-plasma for 90 seconds prior to the metal evaporation. (ii) Au ohmic contacts (80 nm) were deposited using the same procedure as the prior step with the exception of the O-plasma. The Au overlaid the bondpads while making contact with the H-terminated surface. (iii) Isolation regions were patterned and exposed to 100W O-plasma for 90 seconds. This step defined the active regions and electrically isolated the devices. (iv) Two samples were passivated with 25~nm of Al$_2$O$_3$ via atomic layer deposition at 250$^{\circ}$C. The oxide interface provides acceptor states for the 2DHG formation, and also stabilizes the 2DHG over time and over a wide range of temperatures \cite{Kasu_2012,verona_2016}. (v) The passivated samples had the oxides etched at the bond pad regions for probing and wirebonding by submerging the patterned sample in a 20:1~BOE solution for 60 seconds. 

\begin{figure}[t!]
\captionsetup[sub]{labelformat=empty,justification=raggedright,singlelinecheck=false}
    \includegraphics[scale=0.14]{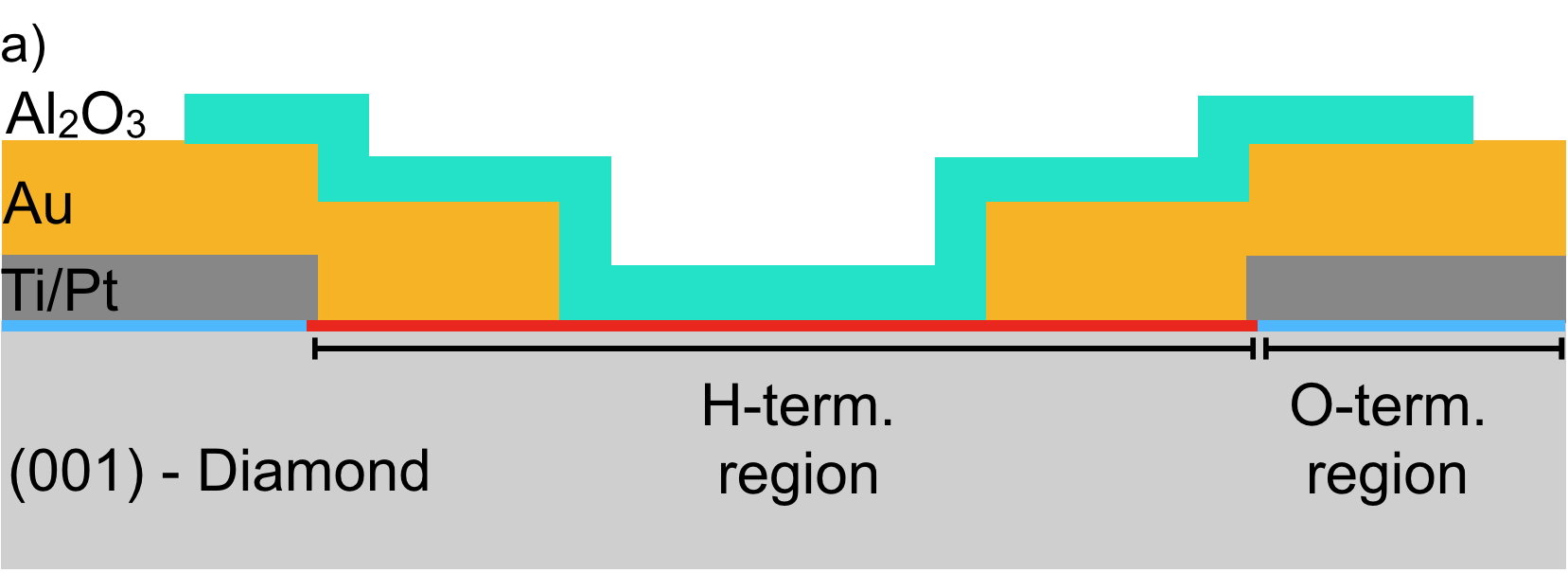}
    \includegraphics[scale=0.14]{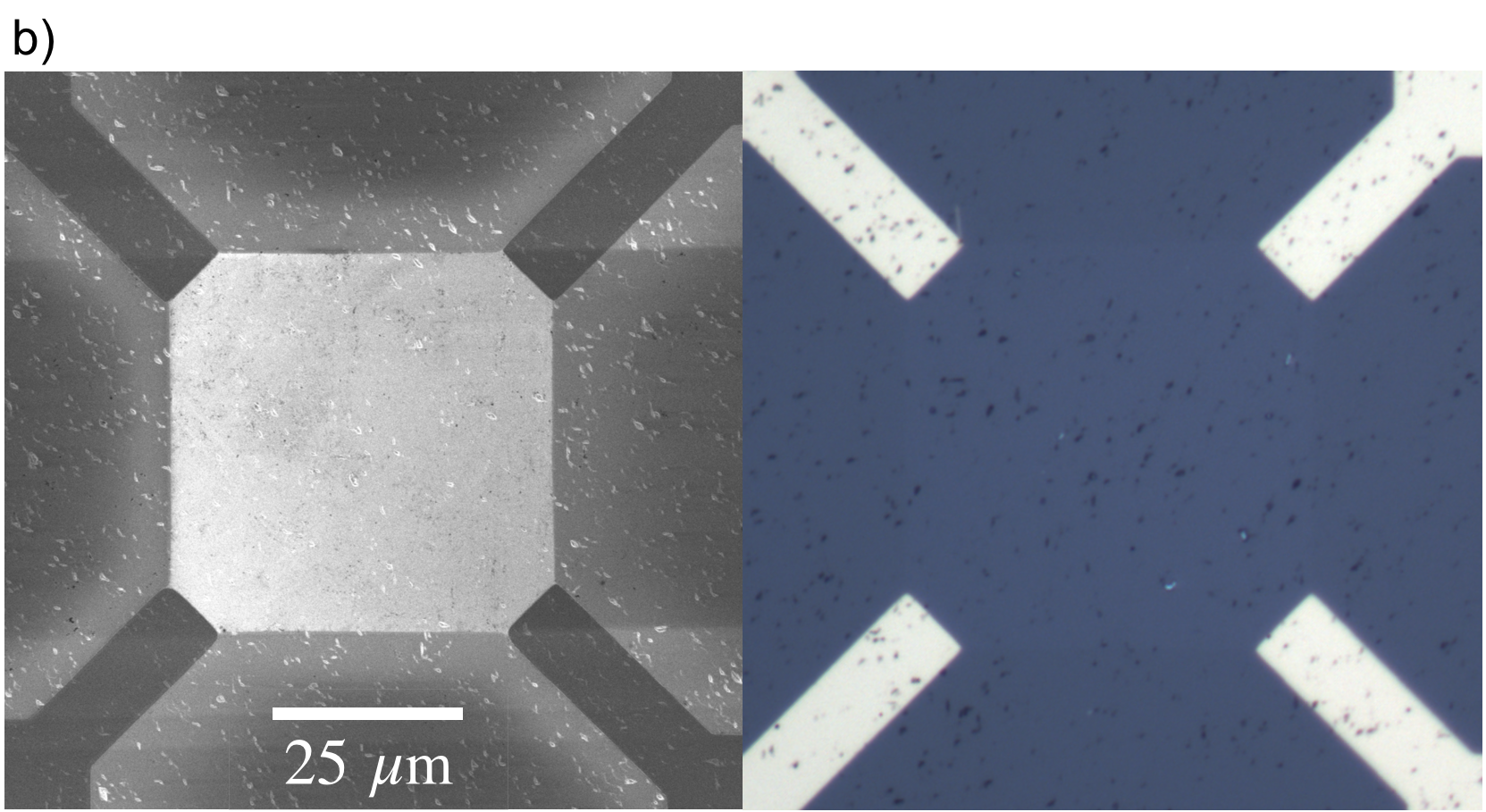}
    \caption{(Color online) (a) Cross-sectional diagram of the H:Diamond Hall-effect structures. (b) Images from an SEM (left) and OM (right) of the fabricated Hall-effect structure. The bright square at the center of the SEM image is the H-terminated active region. The bright and dark spots (shown in the SEM and OM image respectively) are etched pits caused by the H-plasma exposure.}
    \label{fig_xsec_sem}
\end{figure}

After fabrication of the Hall-effect devices, Hall measurements were taken using the Lake Shore 8404 Hall system at temperatures ranging from 25~K to 700~K. Additionally, atomic force microscopy (AFM) measurements were taken in the active regions of the Hall-effect device. A cross-sectional diagram, and images from a scanning electron microscope (SEM) and optical microscope (OM) of the final device are shown in Fig.~\ref{fig_xsec_sem}.

\section{Model and Scattering Mechanisms}

Four scattering mechanisms have been considered in this model. For 2DHGs in H:Diamond, the hole mobility is only limited by phonons at high temperatures. In the low-to-intermediate regime, however, the mobility is theorized to be mostly limited by the coulombic interactions between the 2DHG and localized fields in-plane, such as those induced by ionized surface acceptors. In this section we present the modeling framework for the hole mobility, which consists of calculations of the Fermi energies, probability densities, and relaxation times for each scattering mechanism.

\subsection{Multi-band treatment}
Determining the average hole relaxation time requires calculating the Fermi level with respect to the valence band maximum (VBM), which is unique to each valence band and confined subbands. Moreover, since the scattering matrix elements are usually functions of effective masses, the calculated relaxation times will also be unique to each band. Therefore, in this calculation, three single-band effective mass Schr{\"o}dinger equations are solved and coupled to the Poisson equation. This is performed for each of the heavy hole, light hole, and split-off valence bands (herein denoted by HH, LH, and SO). This calculation was performed using a Schr{\"o}dinger/Poisson solver (nextnano$^3$ software) \cite{nextnano1}, as it already been used in other works for the same 2DHG H:Diamond technology \cite{dia_nn1,dia_nn2}.  To induce a confined accumulation of holes at the surface, a negative interface sheet density was imposed as a boundary condition at the surface. At a given temperature, the negative charge density was modulated until the hole density matched the sheet density extracted from Hall measurements. Finally, a Neumman boundary condition ($\partial \varphi/\partial z = 0$) was set at 500~nm from the surface.

Once at the desired conditions, the Fermi levels for each band were extracted, given by 

\begin{gather}\label{eq:EV_EF}
    \mathcal{E}_{F,{j\ell}} = \mathcal{E}^{VBM}_{j\ell} - \mathcal{E}_F,
\end{gather}

\begin{figure}[b!]
{\includegraphics[scale=0.57,trim={0 0 0.1cm 0},clip]{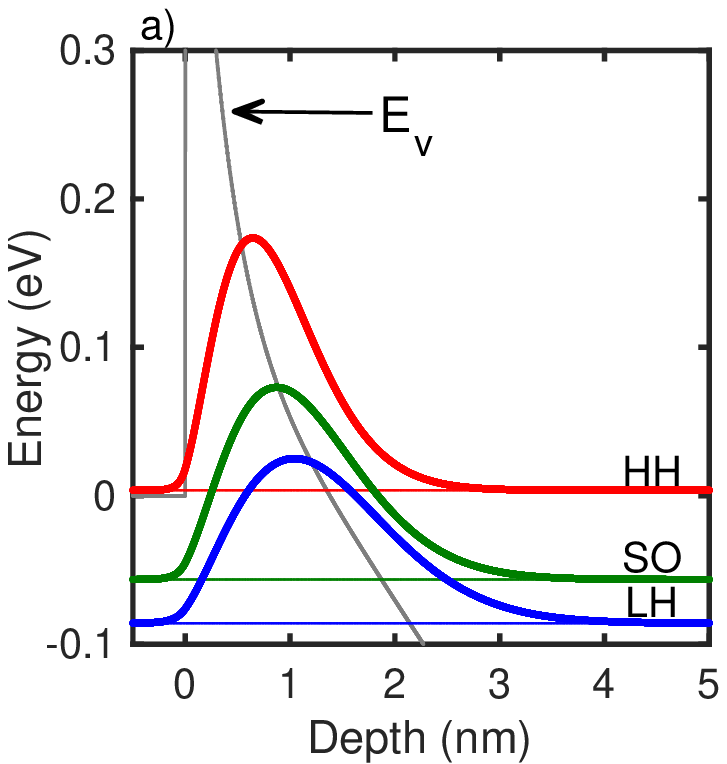}}
{\includegraphics[scale=0.57,trim={0.2cm 0 0.5cm 0},clip]{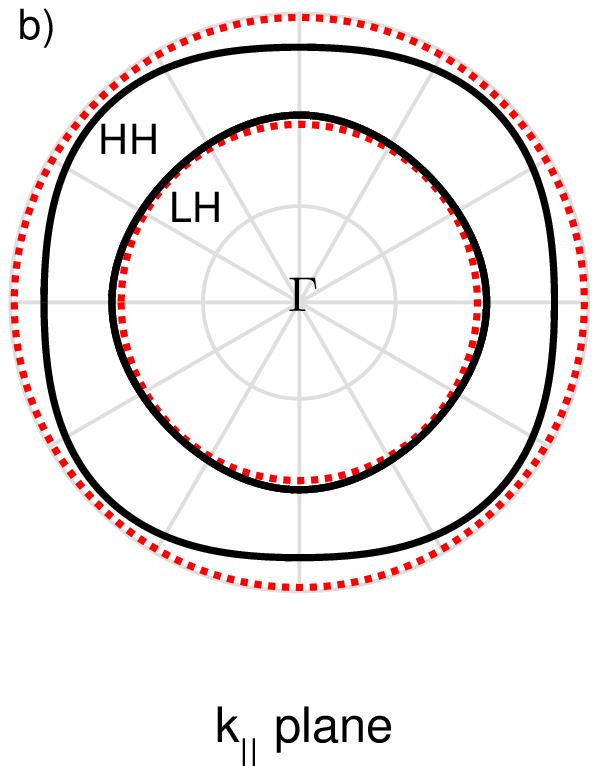}}
\caption{(Color online) (a) Ground state energy ($\ell=1$) of HH, LH, and SO bands at the diamond surface, where the reference energy is the Fermi level at $0$ eV. Superimposed are the hole probability densities for each band. (b) Constant energy surface on a $\mathbf{k}_{||}$ plane of the diamond HH and LH bands. The solid black lines are calculated using the Luttinger parameters by Naka \textit{et al}. \cite{NAKA}. The dashed red lines are a parabolic approximation using the angle-averaged masses.}
\label{fig_psi2_kplane}
\end{figure}

where $\mathcal{E}^{VBM}_{j\ell}$ is the VBM of band $j$ (HH, LH, and SO) and subband $\ell$ (1, 2, and 3), and $\mathcal{E}_F$ is the global Fermi level. The confined hole wave functions out-of-plane $\psi_{j\ell}(z)$ are used for the scattering matrix element calculations. Further, since the relative hole occupation $\rho_{2D}^{j\ell} \propto \left|\psi_{j\ell}(z)\right|^2$, we can justify simplifying our calculations by ignoring higher subbands with a low occupation number. Given the p-like orbital degeneracy of the valence bands, the occupation of holes is dominated by the ground state for each band $j$, even for very high sheet densities. For this reason, only the first subband ($\ell=1$) of each band $j$ is considered for our calculations (Fig.~\ref{fig_psi2_kplane}(a)).

Together with the respective effective masses, this data establishes the starting point for the calculation of the hole relaxation times. 

\subsection{Effective Masses}

The mobility of any crystal structure is in large part influenced by the effective masses of the majority carriers, which therefore ties the diamond band structure into this analysis. As with any semiconductor, the behavior of holes are dictated by the two-fold degenerate HH and LH bands, as well as the SO band separated by $\Delta E_{so}$, located at the $\Gamma$ point in the $E(k)$ dispersion diagram. Typical examples of $\Delta E_{so}$ are 44 and 28 meV for Si and Ge, respectively. Diamond is unusual in this regard, with $\Delta E_{so} \approx 6$~meV~\cite{SO_BAND_exp}. This implies that the hole occupation in the SO band is more probable, hence the importance of the multi-band treatment in our calculations. 

With the exception of holes in the SO band, it is typical that the hole dispersion be highly anisotropic due to warping of the constant energy surfaces of the HH and LH bands. Via the $\mathbf{k}\cdot\mathbf{p}$ perturbation scheme, this dispersion can be analytically expressed as $E(\mathbf{k})_{HH,LH} = A \mathbf{k}^2 \pm [B^2 \mathbf{k}^4 + C^2(k_x^2k_y^2+k_y^2k_z^2+k_z^2k_x^2)^{1/2}$, where $\mathbf{k}~=~\langle k_x,k_y,k_z\rangle$ is the wave vector, and the constants $A$, $B$, and $C$ are functions of the Luttinger parameters determined experimentally~\cite{EK_DISP}. 

Unfortunately, literature on the hole effective masses appears to be inconclusive, given the wide range of values reported \cite{meff_old1,meff_old2}. For our calculations, only the most recent studies are considered. In one study, Y.~Takahide \textit{et al}. measured a range of effective masses through Shubnikov-de Haas oscillations on hydrogen-terminated surfaces for magnetic fields perpendicular to (111) plane. The oscillation peaks corresponded to effective masses which the authors grouped into two separate ranges: $m^*/m_0~=~0.17-0.36$ and $0.57-0.78$ \cite{Takahide}. These ranges reasonably encapsulate the masses reported a year prior by Naka \textit{et al}. using cyclotron resonance experiments~\cite{NAKA}. The latter study provides the most recently obtained Luttinger parameters of $\gamma_1=2.67$, $\gamma_2=-0.403$, and $\gamma_3=0.680$, which in turn yield $2.67$, $-0.8$, and $1.9049$ for constants $A$, $B$, and $C$, respectively. Cross-sections of the constant energy surfaces for HH and LH bands are plotted in Fig.~\ref{fig_psi2_kplane}(b) about the $\Gamma$ point. This plot shows that, although warping of the valence bands is visible, it is reasonable for our purposes to treat dispersion as parabolic (i.e. $E(k)=\hbar^2 k^2/(2m^*)$). Thus, using the data by N. Naka, we use the angle-averaged hole masses for the in-plane effective masses in our scattering model. For the out-of-plane calculations performed by the Schr{\"o}dinger/Poisson solver, the [001] effective masses are used. These masses are listed in Table \ref{table:1}.

\begin{table}[t!]
 \begin{tabular}{ | m{4cm} | c | c | }
 \hline\hline
 Valence Band & Angle-averaged & [001] direction \\ [0.5ex] 
 \hline
 Heavy hole, j=HH & 0.667 & 0.540 \\ 
 Light hole, j=LH & 0.260 & 0.288 \\
 Split-off, j=SO & 0.375 & 0.375 \\
 \hline\hline
\end{tabular}
\caption{Effective masses used for our calculations. Values are in free-electron mass units \cite{NAKA}.}
\label{table:1}
\end{table}

\subsection{Formulation of Scattering Rates}

Devices based on two-dimensional electron conduction have been a subject of extensive research for the past decades, and as such, two-dimensional carrier scattering models have been developed. In this model, the 2D holes are characterized by a plane wave along the diamond surface ($\mathbf{r}$-plane), and a quantized wave perpendicular to the surface ($z$-axis). Thus the incident and final states, expressed as plane waves, are written as ${\Psi_i =A^{-1/2} \psi(z) \text{exp}(i \mathbf{k}\cdot\mathbf{r})}$ and ${\Psi_f =A^{-1/2} \psi(z) \text{exp}(i \mathbf{k}' \cdot\mathbf{r})}$, where $\mathbf{k}$ and $\mathbf{k}'$ are the initial and final hole wave vectors, respectively, and $\psi(z)$ is the out-of-plane wave function determined using the Schr{\"o}dinger/Poisson solver. The factor $A$ is the 2D normalization constant converting the scattering rate per unit area (also denoted by $L^2$). The two-dimensional form of scattering rate is expressed by integrating over all possible final states $\mathbf{k}'$ of the scattering matrix $M_i(\mathbf{k},\mathbf{k})$,


\begin{gather}\label{eq:scat1}
    \Gamma_{\mathbf{k}',\mathbf{k}}^i = \frac{2\pi}{\hbar} \frac{L^2}{(2\pi)^2} \int d^2\mathbf{k}'
    |M_i(\mathbf{k}',\mathbf{k})|^2 \delta[\mathcal{E}_{\mathbf{k}'}-\mathcal{E}_{\mathbf{k}}],
\end{gather}




where $i$ denotes the scattering mechanism, and the delta function ensures the conservation of energy. The mobility is determined by the transport lifetime (or relaxation time) $\tau_{rt}$, which is a function of the net scattering rates $\Gamma^i_{\mathbf{k}',\mathbf{k}}$ and also a function of the scattering angle between vectors $\mathbf{k}$ and $\mathbf{k}'$, denoted by $\theta$. Via Boltzmann transport equation and the principle of detailed balance, the angle dependence is introduced by the factor ($1-\text{cos}(\theta)$), which is intuitive since a scattering angle of $180^{\circ}$ minimizes the transport lifetime, while an angle of $0^{\circ}$ is not treated as a scattering event. We can write the relaxation rate in terms of the displacement vector $\mathbf{q}=\mathbf{k}'-\mathbf{k}$,

\begin{gather}\label{eq:scat2}
    \frac{1}{\tau_i (k)} = \frac{2\pi}{\hbar} \frac{L^2}{(2\pi)^2} \int d^2\mathbf{q}
    |M_i(\mathbf{q})|^2 (1-\text{cos}(\theta)) \delta[\mathcal{E}_{\text{k}'}-\mathcal{E}_\text{k}].
\end{gather}

Presuming each scattering mechanism $i$ is independent, the total relaxation time $\tau_{tr}$ is given by

\begin{gather}\label{eq:tau_tr}
    \frac{1}{\tau_{tr} (k)} = \sum_i \frac{1}{\tau_i(k)}.
\end{gather}

These relaxation times are numerically calculated and averaged according to the Fermi statistics,

\begin{gather}\label{eq:tau_avg}
    \left. \langle \tau_{tr}\rangle_j = \sum_\text{k}\mathcal{E}_{\text{k}} \tau_{tr}^j(k)
    \left(\frac{\partial f(\mathcal{E}_{\text{k}})_j}{{\partial\mathcal{E}_{\text{k}}}}\right) \middle/ \sum_\text{k}\mathcal{E}_{\text{k}}\left(\frac{\partial f(\mathcal{E}_{\text{k}})_j}{{\partial\mathcal{E}_{\text{k}}}}\right) \right. .
\end{gather}

Here the subscript $j$ was introduced to signify that the relaxation times are unique to each band $j$, each of which has an effective mass $m_j^*$, carrier density $\rho_{2D}^j$, and Fermi energy from Eq. \eqref{eq:EV_EF}. The averaged relaxation time is used to deduce the hole mobility, obtained using the widely used relation,

\begin{gather}\label{eq:mu_j}
    \mu_j = \frac{e}{m^*_j} \langle \tau_{tr} \rangle_j.
\end{gather}

Finally, the total mobility can be determined by weight-averaging each band, given by

\begin{gather}\label{eq:muH}
    \mu_H = \frac{\sum_j \mu_j \rho_{2D}^{j}}{\sum_j \rho_{2D}^{j}}.
\end{gather}

The final result in Eq.~\eqref{eq:muH} links this theoretical framework with the measured quantity obtained via Hall measurements. The scattering mechanisms modeled by the matrix elements in Eq.~\eqref{eq:scat2}, as well as the relaxation averaging of Eq.~\eqref{eq:tau_avg}, result in $\mu_H$ that is a function of temperature, sheet carrier density, impurity density, and other material properties. This modeling framework will thus provide us with a thorough understanding of the limitations to the hole conductivity of 2D hole gases in H:Diamond.

\subsection{Surface Impurities (SI)}

A major consequence of the charge transfer phenomenon is that the 2DHG is compensated by negatively charged acceptor states, which can be provided by air-adsorbates or oxide films. The sheet separation of the hole gas and negative compensating charge is on the order of angstroms. Thus, the induced Coulombic forces perturb the 2D potential well, which significantly degrades the hole mobility. Such 2D carrier channels in other material stacks, such as the 2DEG in remotely-doped AlGaAs/GaAs heterostructures, are relatively distant from the charged donors, which allows for electron mobilities as high as $10^4~\text{cm}^2$/(V$\cdot$s). However, even for these structures, the mobility can be limited by these remote impurities, especially at low temperatures. Hence, this scattering mechanism has been modeled for 2D carriers, and is adopted herein \cite{davies_1997}. Moreover, disorder related to the \mbox{C-H} dipoles (e.g., incomplete hydrogen termination \cite{Nebel_2002,Garrido_2005}, non-homogeneous acceptor distribution~\cite{SATO2012}, variation in surface reconstruction \cite{KAWARADA1996}, the existence of oxygen-related catalysts \cite{Riedel_2004}, etc.) may interfere with the conductivity of holes. Together with the negatively charged acceptor states, these field-inducing phenomena distort the band structure and thus act as scattering centers. We will unravel this further in the discussion and results section. Here we denote the scattering by the negatively charged surface acceptors as type (i), and the scattering induced by the C-H disorder as type (ii). The matrix element is expressed as

\begin{align}\label{eq:M_SI1}
    M_{si}(\mathbf{q}) &= \int_0^\infty \left|\psi(z)\right|^2 dz \int V(\mathbf{r},z) \text{exp}(i\mathbf{q}\cdot\mathbf{r}) d^2 \mathbf{r}, \nonumber\\ 
    &= \int_0^\infty \left|\psi(z)\right|^2 \mathcal{V}(q,z) dz ,
\end{align}
    
where $\mathcal{V}(q,z)$ is the Fourier transform of the potential form of a charged impurity. Following Ref. \cite{davies_1997}, properly taking charged screening into account gives us

\begin{gather}\label{eq:M_SI2}
    \mathcal{V}(q,z) = \frac{Z e^2}{2\varepsilon_0\epsilon(q)} \frac{\text{exp}(-q(z+|d|))}{q},
\end{gather}

where $Z$ is the electronic charge number and $\epsilon(q)$ is the dielectric constant defined by

\begin{gather}\label{eq:M_SI_eps_q}
    \epsilon(q) = \varepsilon_s \left(1 + \frac{q_{TF}F(q)}{q} \right).
\end{gather}

Here, screening is treated via the 2D Thomas-Fermi wave vector ~$q_{TF}=m^*_{dos} e^2/(2\pi\varepsilon_0\varepsilon_s\hbar^2)$~, and $F(q)$ is a form factor defined by

\begin{gather}\label{eq:M_SI_FF}
    F(q) = \int dz \int dz' |\psi(z)|^2|\psi(z')|^2\text{exp}\left(-q|z-z'|\right).
\end{gather}

With the wave functions (confined along $z$) and the wave vectors treated parabolically in two-dimensions, the scattering rate can be expressed as

\begin{multline}\label{eq:M_SI3}
    \frac{1}{\tau_{si}^{(i),(ii)}} = \frac{(Z^2 N_{si})^{(i),(ii)} m^*_{dos}}{2\pi\hbar^3k^3} \left(\frac{e^2}{2\varepsilon_0\varepsilon_s}\right)^2 \int_0^{\infty}dz\left|\psi(z)\right|^2 \\ \times \int_0^{2k}\frac{\text{exp}(-2q(z+|d|))}{(q+F(q)q_{TF})^2} \frac{q^2dq}{\sqrt{1-(q/2k)^2}},
\end{multline}

where $(Z^2 N_{si})^{(i)}$ and $(Z^2 N_{si})^{(ii)}$ are the fitting parameters for SI scattering of types (i) and (ii), respectively. Throughout the text, $Z$ is absorbed into the fitting parameter for type (ii) scattering since the nature of the induced fields is uncertain. Thus we define \mbox{$\mathcal{N}_{si}^{(ii)}=(Z^2 N_{si})^{(ii)}$}. For type (i) scattering, however, each ionized surface acceptor is presumed to have a charge of unity. Thus we set $(Z^2 N_{si})^{(i)} = N_{si}^{(i)}$. 

\subsection{Surface Roughness (SR)}

Roughness in the form of spatial fluctuations at the H:Diamond surface may be induced via diamond growth, exposure to hydrogen plasma, or during the fabrication process. Hence, the fluctuations produce localized potentials randomly distributed along the plane, which act as scattering centers for holes. If the fluctuations are on the order of carrier wavelengths, then scattering can be significant. We denote the average out-of-plane fluctuations as $\Delta$ (i.e., root-mean-squared (RMS) roughness height) and the average in-plane separation of these fluctuations as $\Lambda$ (i.e., correlation length). These roughness variables are expressed by a Gaussian distribution as ${\langle \Delta(\mathbf{r})\Delta(\mathbf{r}') \rangle = \Delta^2\text{exp}\left[(\mathbf{r}-\mathbf{r}')^2/\Lambda^2\right]}$. The formalism by Ando \textit{et al}. \cite{ando_1982} is adopted here for H:Diamond, which expresses the scattering matrix element as

\begin{gather}\label{eq:M_SR1}
    |M_{sr}(\mathbf{q})|^2 = \frac{e^4 \rho_{2D}^2}{4\varepsilon^2}\frac{\pi \Delta^2\Lambda^2}{L^2}\text{exp}\left(\frac{-q^2\Lambda^2}{4}\right),
\end{gather}

where the Fourier transform of the $\mathbf{r}$-Gaussian distribution $\langle | \Delta(\mathbf{q})|^2 \rangle$ was used for the matrix element. Here we presume that the sheet hole density $\rho_{2D}$ is the only form of charge and ignore other variants (e.g., space charge density). With the substitution of Eq.~\eqref{eq:M_SR1} into the 2D transport lifetime expression and integrating over the wave vector plane, the final form is

\begin{gather}\label{eq:M_SR2}
    \frac{1}{\tau_{sr}} = \frac{\pi m^*_{dos} \Delta^2 \Lambda^2 e^4 \rho_{2D}^2}{\hbar^3 (\varepsilon_0 \epsilon(q))^2} \text{exp}\left(-\frac{q^2\Lambda^2}{4}\right),
\end{gather}

Note from Eq.~\eqref{eq:M_SR2} that the scattering rate increases with the square of $\rho_{2D}$, and may thus be insignificant at low sheet densities. 

\subsection{Non-Polar Optical Phonons (NOP)}
Scattering of carriers by phonons dominate at high temperatures, which is an intrinsic phenomenon in all materials. Thus, hole-phonon interactions are dependent on the physical parameters of the material, such as the effective mass, material density, and (in the case of carriers confined to a 2D plane) the $z$-plane probability density of the 2DHG. It is for this reason that in the limit of higher temperatures, carrier-phonon interactions are the insurmountable limiting factor of carrier mobilities. In this section, we define the relaxation time for holes interactions with non-polar optical phonon (NOP), which exhibits a steep slope with respect to temperature.

This scattering matrix is commonly defined as the product of the deformation potential $D_{op}$ and the optical phonon displacement vector $\mathbf{u}_{op}$, expressed as ${|M_i(\mathbf{q})| = D_{op} \cdot \mathbf{u}_{op}}$. The displacement vector, derived in Ref. \cite{Hamaguchi2017}, yields the scattering matrix

\begin{gather}\label{eq:M_NOP1}
    |M_{nop}(\mathbf{q})|^2 = \frac{D_{op}^2\hbar}{2\rho L^3 \omega_0} \times \left(n(\omega_0) + \frac{1}{2} \pm \frac{1}{2}\right),
\end{gather}

where $\rho$ is the material density of diamond, $n(\omega_0)$ is the phonon occupation factor, and $\hbar\omega_0$ is the NOP energy, which is assumed to be dispersion-less and thus independent of $\mathbf{q}$. Due to confinement along the $z$-direction, carriers are restricted along the $\mathbf{r}_{||}$ plane, while phonons are treated in three dimensions ($\mathbf{q}^2 + q^2_z$). Hence, the three-dimensional form of Eq.~\eqref{eq:scat2} is used and quantized along the $q_z$ direction. The final expression yields

\begin{multline}\label{eq:M_NOP2}
    \frac{1}{\tau_{nop}} = \int |I(q_z)|^2 dq_z \cdot \frac{m_{dos}^* D_{nop}^2}{4\pi\rho\hbar^2\omega_0} \\ \times \left(n_0(\omega_0) + \frac{1}{2} \pm \frac{1}{2}\right) \Theta(\mathcal{E}_k \mp \hbar\omega_0),
\end{multline}

where  the overlap integral is defined by  ${|I(q_z)|=\int \left|\psi(z)\right|^2\text{exp}(iq_z z)dz}$, and $\Theta(x)$ is the step function which is unity when $x\geq0$ and zero otherwise. Here we also recognized that the integration over ${\delta[\mathcal{E}_{\mathbf{k}'}-\mathcal{E}_{\mathbf{k}}]/(2\pi)^2}$ is the definition of the 2D density of states $m^*_{dos} / \pi \hbar^2$.

\subsection{Acoustic Phonons (AP)}\label{eq:M_AP1}
At low to intermediate temperatures, acoustic phonons are the most dominant species of electron-phonon scattering. As with NOP scattering, the potential $D_{ac}$ and the acoustic phonon displacement vector $\mathbf{u}_{ap}$ define the scattering matrix as

\begin{align}\label{eq:M_AP2}
    |M_{ap}(\mathbf{q})|^2 &= \frac{q^2 D_{ap}^2\hbar}{2\rho L^3 \omega_q} \times \left(n(\omega_q) + \frac{1}{2} \pm \frac{1}{2}\right), \nonumber\\ 
    &= \frac{D_{ap}^2 k_B T}{2\rho L^3 v_s^2},
\end{align}

where $v_s$ is the longitudinal sound velocity. Here we invoked the equipartition theorem, where $\hbar\omega_q \ll k_B T$, therefore the phonon occupation number ${n(\omega_q) = 1/(\text{exp}(\hbar\omega_q/k_B T) - 1))} \gg 1$. Hence we can say that ${n(\omega_q) \approx n(\omega_q)+1 \approx k_B T / \hbar \omega_q}$. We also treat the acoustic dispersion relation as linear, i.e. $\omega_q \approx v_s q$.  Treating the integration similarly as in Eqs.~\eqref{eq:M_NOP1}--\eqref{eq:M_NOP2} yields

\begin{gather}\label{eq:M_AP3}
    \frac{1}{\tau_{ap}} = \int |I(q_z)|^2 dq_z \cdot \frac{m^*_{dos} k_B T D_{ap}^2}{\pi\hbar^3\rho v_s^2}.
\end{gather}

\section{Results and Discussion}
In this section we analyze the multiple scattering mechanisms associated with hole transport in H:Diamond using the experimental data of the fabricated Hall-effect structures, as well as work reported previously in literature. We begin by comparing our model to a model previously reported by Y. Li \textit{et al}.~\cite{YLIetal} using experimental data by H. Kasu \textit{et al}.~\cite{Kasu_2012}, which was reported to have a high sheet density of $\sim~4\times10^{13}$ cm$^{-2}$. The previous model primarily used SI and SR scattering for fitting to the data at low-to-intermediate temperatures. Here we repeat this fitting using our model, which allows us to test our calculations to the limit of higher sheet densities. This starting point will subsequently illuminate the shortcomings of only considering type (i) SI and SR scattering, which turn out to be insufficient for lower sheet densities. A complete version of the model, which considers type (ii) SI scattering, will be compared to the Hall measurements on the fabricated devices of this work, where sheet densities are as low as $\sim 2\times 10^{12}$ cm$^{-2}$. 

\begin{figure}[b!]
\includegraphics[scale=0.65,trim={0.25cm 0 0.1cm 0.25cm},clip]{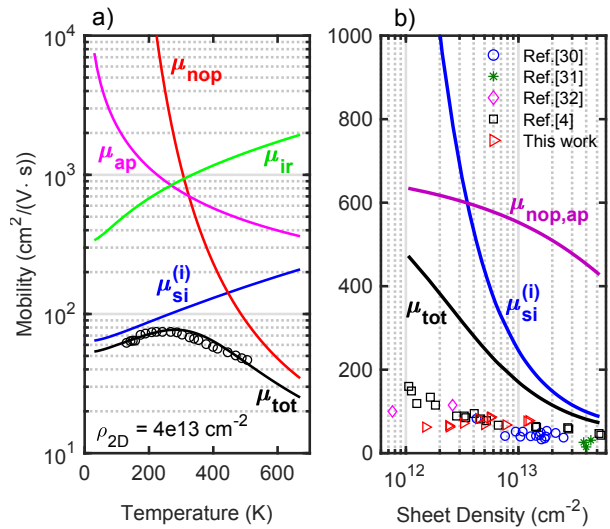}
\caption{(Color online) (a) Calculated Hall mobility as a function of temperature. The data points are reported by H. Kasu \textit{et al}. \cite{Kasu_2012}, where the sheet density was approximately $4\times10^{13}$~cm$^{-2}$. Surface roughness was fitted with parameters $\Delta=1.2$~nm and $\Lambda=5$~nm. (b) Calculated mobilities at T = 300~K with scattering by phonons and SI of type (i) are included. As shown, there is poor agreement with the total mobility and the multiple data points at low sheet density~\cite{Hirama_2008,CRAWFORD2018,OING2019,verona_2016}.}
\label{mutot_kasu_muSI_expPoints}
\end{figure}

The fitting of our calculations to the data by H. Kasu \textit{et al}. is presented in Fig.~\ref{mutot_kasu_muSI_expPoints}(a). The material parameters used are listed in Table~\ref{table:2}. In the mobility model by Y. Li \textit{et al}., approximations such as the 2D Fermi wave vector ($k_F=\sqrt{2\pi\rho_{2D}}$) and a single equivalent isotropic valley model were used. This therefore yielded temperature independent functions for SI and SR scattering, as well as distinct fitting parameters. To performed this calculation using our multi-band treatment and averaging over energy (Eq. (\ref{eq:tau_avg})), we select the same value for the RMS roughness height as Y. Li \textit{et al}., $\Delta = 1.2$~nm, which is a reasonable value taken from Ref.~\cite{Hirama_2008}. The correlation length $\Lambda$ was fitted to be $5$~nm. For NOP scattering, the coupling constant $D_{nop}$ was fitted to be $1.4~\times~10^{10}\; \text{eV/cm}$. As for type (i) SI scattering, it is presumed that the sheet separation of the charged surface acceptors and the 2DHG is the summation of the C-H dipole bond length ($\sim 1.1 \mathrm{\AA}$, \cite{Liu_Fengbin_2015}) and half the thickness of the negatively charged acceptors ($\sim 2 \mathrm{\AA}$, \cite{Tsugawa_2001}), which gives us $d = 2.1~\mathrm{\AA}$. Finally, the negative surface acceptor density was presumed to exactly balance the positive sheet density, giving $N_{si}^{(i)} =\rho_{2D}$.

\begin{table}[t!]
 \begin{tabular}{ | m{4cm} | c | c | }
 \hline\hline
 Parameter & Symbol (units) & Value [Ref.] \\ [0.5ex] 
 \hline
 NOP Deformation potential & $D_{op}$ (eV/cm) & 1.4e10 \\ 
 AP Deformation potential & $D_{ap}$ (eV) & 8 \cite{Pernot_2010} \\
 LO-phonon energy & $\hbar\omega_0$ (meV) & 165 \cite{Pernot_2010}\\
 Material density & $\rho$ (kg/m$^3$) & 3515 \\
 Sound velocity & $v_s$ (m/s) & 17536 \\
 Dielectric constant & $\varepsilon_s$ ($\varepsilon_0$) & 5.7\\
 Surface acceptor separation& d ($\mathrm{\AA}$) & 2.1 \\
 \hline\hline
\end{tabular}
\caption{Material parameters used in the 2DHG H:Diamond scattering calculations.}
\label{table:2}
\end{table}

Fig.~\ref{mutot_kasu_muSI_expPoints}(a) shows that SI scattering by negatively charged acceptors (i.e., type (i)) is the dominant mechanism, particularly at low to intermediate temperatures, which is attributed to the high sheet density of $\sim 4\times10^{13}\;\text{cm}^{-2}$. It is important to note the slight decrease in $\mu_{tot}$ at lower temperatures. Since ionized impurity scattering is much higher near the valence band edge (i.e., the top of the valence band), the Fermi energy averaging of holes (Eq.~(\ref{eq:tau_avg})) is necessary to capture this behavior. As shown in Fig.~\ref{mutot_kasu_muSI_expPoints}(b), the type (i) dominance is further evident above $1\times10^{13}$~cm$^{-2}$, where $\mu_{si}^{(i)}$ drops to commonly measured mobility values. However, as given by the factors in Eqs. (\ref{eq:M_SI3}) and (\ref{eq:M_SR2}), the SI and SR scattering rates increase with $\rho_{2D} (=N_{si}^{(i)})$ and $\rho_{2D}^2$, respectively. Thus, at $\rho_{2D} \ll 1\times10^{13}$~cm$^{-2}$ and at 300~K, this modeling framework predicts a total hole mobility that is limited by phonons, which is significantly higher than what is measured experimentally, as shown in Fig.~\ref{mutot_kasu_muSI_expPoints}(b). This inaccuracy is further evident at lower temperatures, where scattering by phonons becomes negligible. Thus, it is clear that an additional scattering mechanism is required to explain this behavior. 

\begin{figure}[t!]
\includegraphics[scale=0.60,trim={0cm 2.5cm 0cm 0cm},clip]{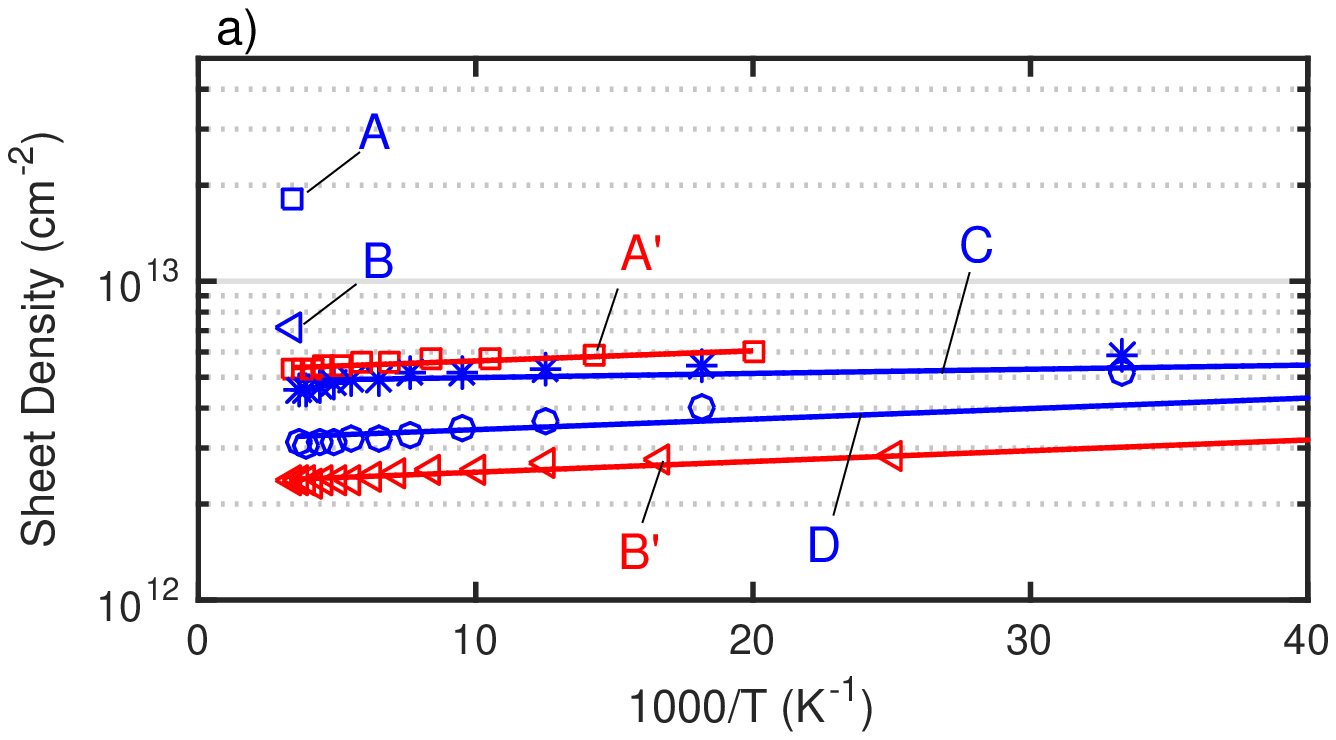}
\includegraphics[scale=0.64,trim={0.7cm 0 1cm 0},clip]{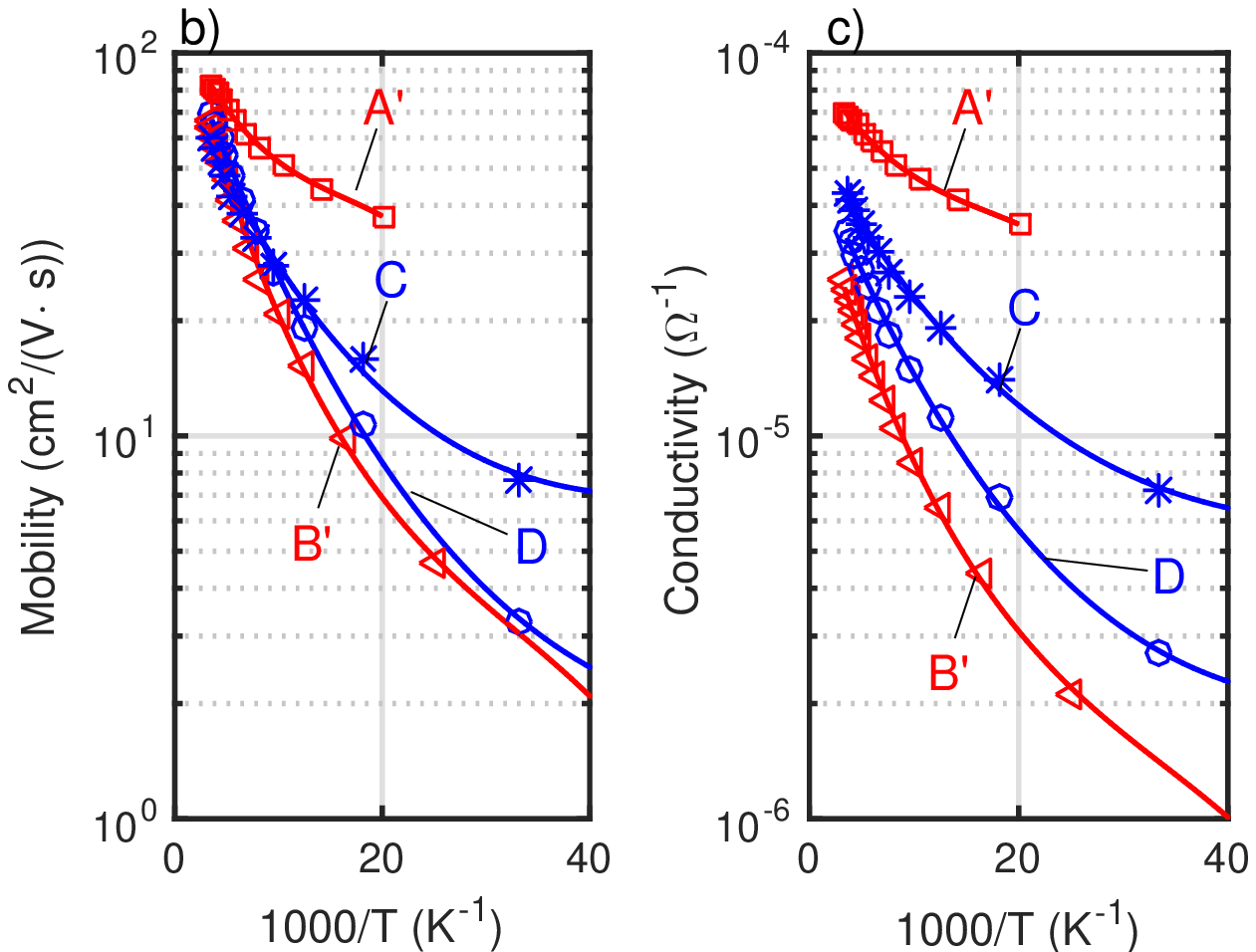}
\caption{(Color online) Hall measurements of the four fabricated samples as a function of inverse temperature. (a) Measured sheet densities for samples A, B, C, and D. We use the prime notation (i.e., A' and B') to denote passivation with Al$_2$O$_3$. The corresponding Hall mobility (b) and conductivity (c) decline at a rate that is dependent on the sheet density (a).}
\label{rho_mu_sigE_sc14_sc19}
\end{figure}

To explore this further, Hall measurements were taken on the fabricated Hall-effect devices. Prior to passivation with Al$_2$O$_3$, Hall measurements of the samples, denoted as A, B, C, and D, were performed after several days of being air-exposed. Afterwards, samples A and B were passivated with 25~nm of ALD-Al$_2$O$_3$, which we denote as samples A' and B'. Finally, Hall measurements were performed over the range $\sim$25~K to 300~K. The measurements were taken from RT to low temperatures, and back up to RT, and negligible hysteresis was observed. The results are shown in Fig. \ref{rho_mu_sigE_sc14_sc19}, plotted as a function of inverse temperature.

We first observe that, prior to passivating samples A and B, the sheet densities are significantly higher than their passivated counterparts (A' and B'). As reported in other works, a drop in the sheet density after the deposition of Al$_2$O$_3$ is common \cite{verona_2016,Ren_2018}, which is attributed to the lower density of surface acceptors in Al$_2$O$_3$ in comparison to air-adsorbates. It is also shown in Fig. \ref{rho_mu_sigE_sc14_sc19}(a) that the $\rho_{2D}$ is rather constant as the substrate temperature drops to $<$50~K. One study by C. Nebel \textit{et al}. reported a hole \textquote{freeze-out} with a critical temperature of 70~K on H:Diamond surface, a phenomenon that is explained by a classical mobility-edge model \cite{Nebel_2002}. Carrier \textquote{freeze-out} is observed when the sheet density collapses below a critical temperature. For H:Diamond, this would be attributed to a confinement of holes into so-called \textquote{localized states} existing near the valence band edge, presumed to arise from $\it{short}$-$\it{range}$ potential fluctuations at the surface. Above the critical temperature, holes posses the thermal energy to excite into de-localized energy states which span the plane of the 2D well (i.e., \textquote{extended states}). This state transition allowed the holes to conduct freely.

Such a \textquote{freeze-out} phenomenon, however, was not observed here. Instead, our work is consistent with what was reported by J. Garrido \textit{et al}., whereby the conductive properties of the 2DHG exhibit a temperature-independent $\rho_{2D}$ \cite{Garrido_2005}. Moreover, a thermal activation energy is observed for the mobility and conductivity, as shown in Figs.~$\ref{rho_mu_sigE_sc14_sc19}$(b) and (c), respectively. Specifically, the mobility and conductivity of the samples with lower sheet densities have a higher thermal activation energy (i.e., the decreasing rate is higher). As discussed by J. Garrido \textit{et al}., this behavior can be explained by an early model formulated by E. Arnold, which predicted a similar temperature dependence of the Hall mobility, conductivity, and sheet density in the case of inverted 2D electron channels in Si/SiO$_2$ structures \cite{ARNOLD_1976}. Using semi-classical percolation theory, Arnold explained that electrons conduct in the presence of $\it{long}$-$\it{range}$ potential fluctuations along the conduction band $E_C$, where \textquote{metallic} regions ($E_F > E_C$) coexist with \textquote{insulating} regions ($E_F < E_C$). When the Fermi energy range is narrow ($T\xrightarrow{}0\;K$), electrons percolate around the \textquote{insulating} regions via the \textquote{metallic} network. Thus, since a lower sheet density has a Fermi level that is much closer to $E_C$, the rate at which electrons scatter is enhanced as the Fermi energy range is narrowed. 

\begin{figure}[b!]
\captionsetup[sub]{labelformat=empty,justification=raggedright,singlelinecheck=false}
    \includegraphics[scale=0.15]{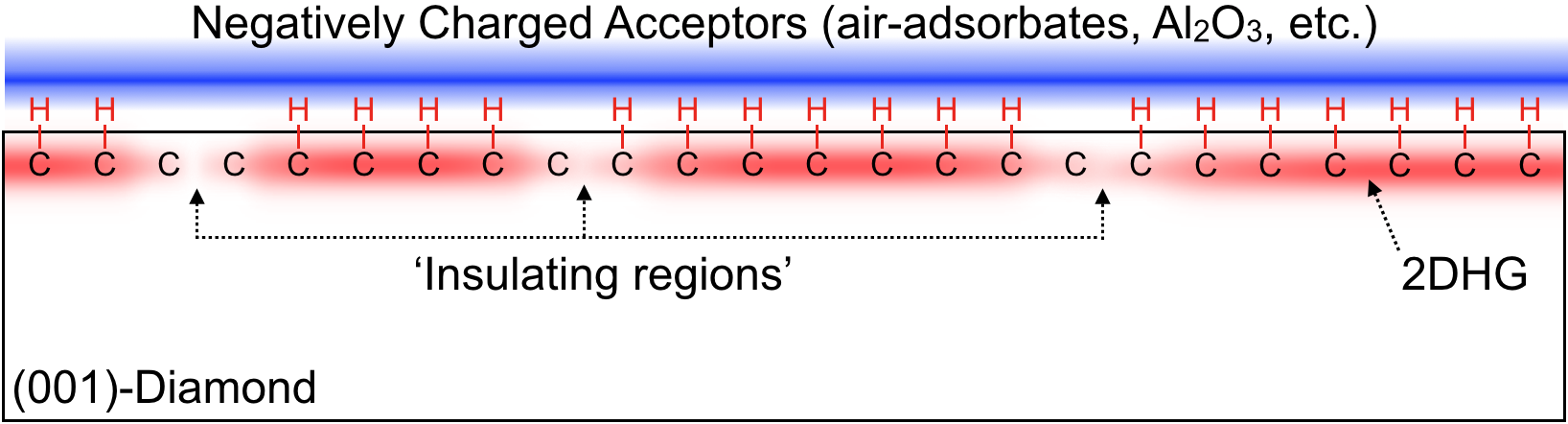}
    \caption{(Color online) Schematic of the insulating regions along the 2DHG caused by incomplete H-termination. Other irregularities related to the C-H surface may also induce this effect.}
    \label{c-h_schematic}
\end{figure}

In light of Arnold's framework, we hypothesize that holes in the 2D well percolate around long-range potential fluctuations induced by the surface impurities of types (i) and (ii). This is evident from the behavior in the mobility and conductivity in Fig.~$\ref{rho_mu_sigE_sc14_sc19}$, where the general trend of increasing activation energy with decreasing sheet density is clearly observed. However, the measurements for samples A' and C exhibit an unusual difference. According to Arnold's framework (subsequently reinforced by J. Garrido \textit{et al}. for H:Diamond), the measured sheet density should be a strong indicator of the activation energy. However, the sheet densities of samples A' and C are very similar, yet yield significantly different activation energies. One explanation that can resolve this inconsistency is to presume that sample A' has a higher periodicity of C-H dipoles at the surface. This presumption is explained by noting the measured sheet density for sample~A' prior to passivation, which was $1.80\times 10^{13}$ cm$^{-2}$. This is four-fold higher than the sheet density of sample~C ($\sim4.50\times 10^{12}$ cm$^{-2}$). As was reported by K. Hirama \textit{et al}., a higher density of C-H dipoles induces a higher hole sheet density \cite{Hirama_2008}. Thus, if we presume that sample C has a lower C-H dipole density as sample A', then under Arnold's framework, holes would \textquote{percolate} around a larger density of \textquote{insulating regions} in sample C than in A' [Fig.~\ref{c-h_schematic}]. The result would thus be a higher activation energy for sample C, despite having a comparable sheet density and Fermi energy as sample~A'.

In the work by K. Hirama \textit{et al}., the reports that a higher C-H dipole density induces a higher sheet density was in the context of out-of-plane orientations of single-crystal diamond, where the carbon density of the restructured surface in the (110) orientation is greater than the (001). Hence, after exposure to a hydrogen plasma, the (110) surface yielded a higher C-H and hole sheet density than the (001). Moreover, this same study showed that a higher CVD temperature induced a higher sheet density, which is likely explained by a more complete H-termination. This was shown directly by T. Ando \textit{et al}., where Fourier-transform infrared spectra exhibited a stronger signal of C-H vibrations for diamond powder that was H-terminated at higher CVD temperatures \cite{Ando_1993}. Finally, other complex surface phenomena, such as a non-homogeneous distribution of surface acceptors~\cite{SATO2012}, the existence of oxygen-related sites~\cite{Riedel_2004}, or a variation of C-H surface reconstruction~\cite{KAWARADA1996}, may contribute to this C-H disorder, and thus strongly influence the behavior observed for the hole mobility.

\begin{figure}[b!]
\includegraphics[scale=0.68,trim={0.1cm 0 0.1cm 0.5cm},clip]{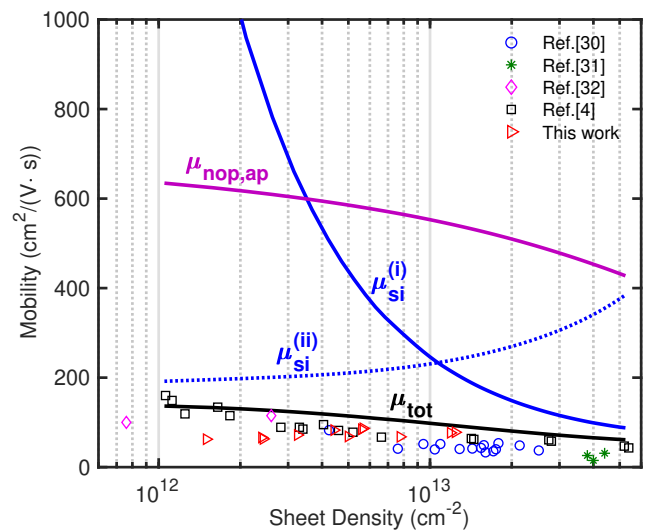}
\caption{(Color online) Measured and calculated Hall mobilities as a function of sheet density at T = 300~K. A value of $\mathcal{N}_{si}^{(ii)} = 5\times10^{12}$~cm$^{-2}$ was arbitrarily selected to demonstrate the trend of type (ii) SI scattering as a function of sheet density. Unlike in Fig.~$\ref{mutot_kasu_muSI_expPoints}$(b), the calculated total mobility is much closer to the experimental values.}
\label{mu_v_expData_wHTD}
\end{figure}

Additionally, such C-H disorder can explain the discrepancy shown in Fig.~$\ref{mutot_kasu_muSI_expPoints}$(b), where the experimental mobility remains relatively stagnant even at low sheet densities. Here we attempt to model this phenomenon, which we denote as SI scattering of type~(ii). Unlike type (i) where $N_{si}^{(i)} = \rho_{2D}$ and $d>0$, type~(ii) SI scattering is related to the surface chemistry of H:Diamond, with $d=0$ and a fitting parameter denoted by $\mathcal{N}_{si}^{(ii)}$. Thus, although $N_{si}^{(i)}$ declines with $\rho_{2D}$, the value of $\mathcal{N}_{si}^{(ii)}$ may remain constant or increase.

Figure~$\ref{mu_v_expData_wHTD}$ is a duplicate of Fig.~$\ref{mutot_kasu_muSI_expPoints}$(b) with $\mu_{si}^{(ii)}$ included. The fitting parameter chosen was $\mathcal{N}_{si}^{(ii)}=5\times10^{12}\;\text{cm}^{-2}$. It is evident here that an increasing sheet density -- and thus Fermi energy -- gives a steady rise in $\mu_{si}^{(ii)}$. This is expected given that SI scattering is more prominent near the top of the valence band. On the other hand, however, $N_{si}^{(i)}$ increases with the sheet density, which reduces $\mu_{si}^{(i)}$. Thus, for the chosen value of $\mathcal{N}_{si}^{(ii)}$, a cross-over point of SI scattering of types (i) and (ii) arises near $1\times10^{13}\;\text{cm}^{-2}$. The result of combining both types SI scattering is to effectively create a mobility \textquote{ceiling} for holes in H:Diamond surfaces, which agrees well with the experimental Hall data from multiple references.

\begin{figure}[t!]
\includegraphics[scale=0.65,trim={0.1cm 0 0.1cm 0},clip]{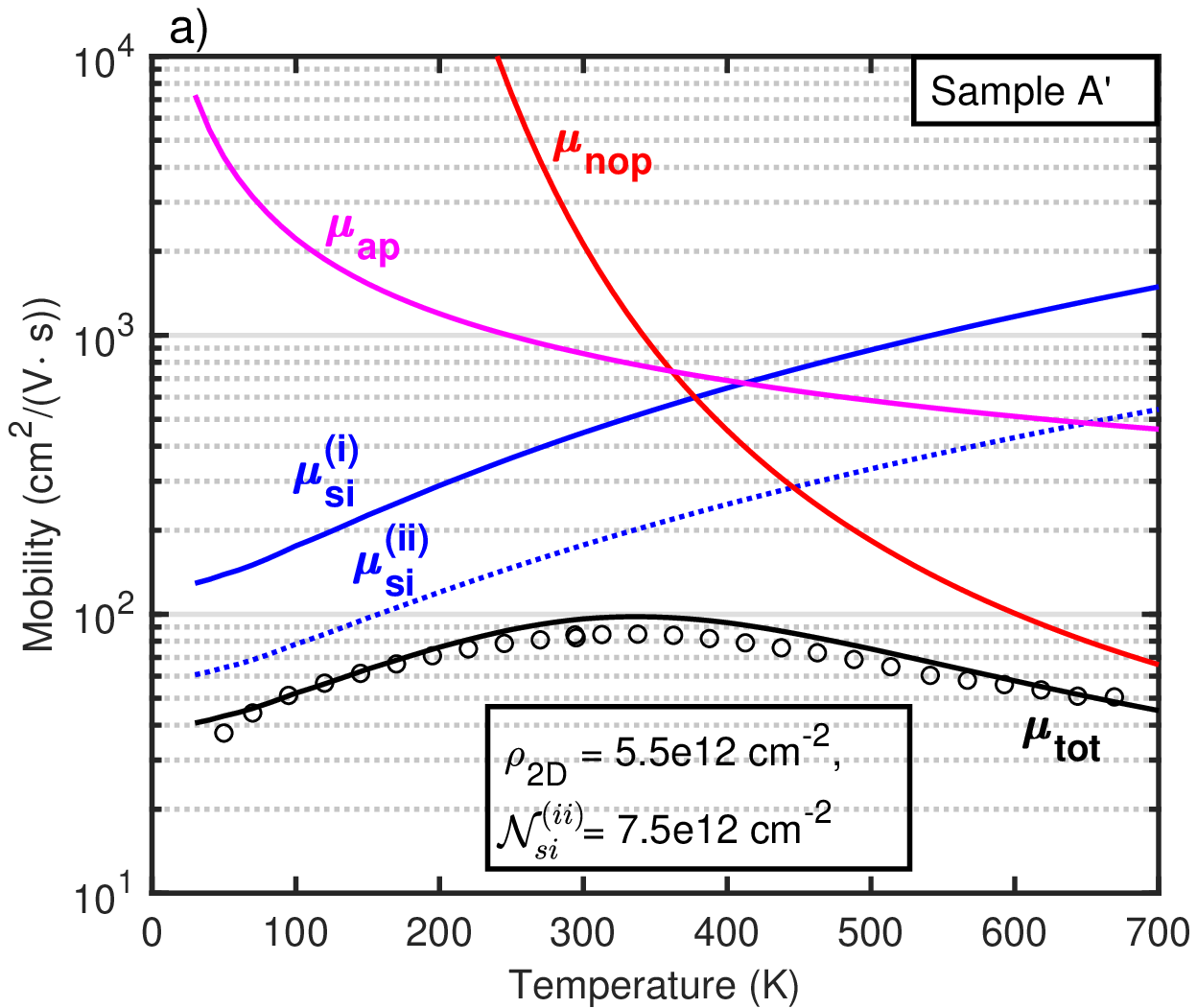}
\includegraphics[scale=0.65,trim={0.1cm 0 0.1cm 0},clip]{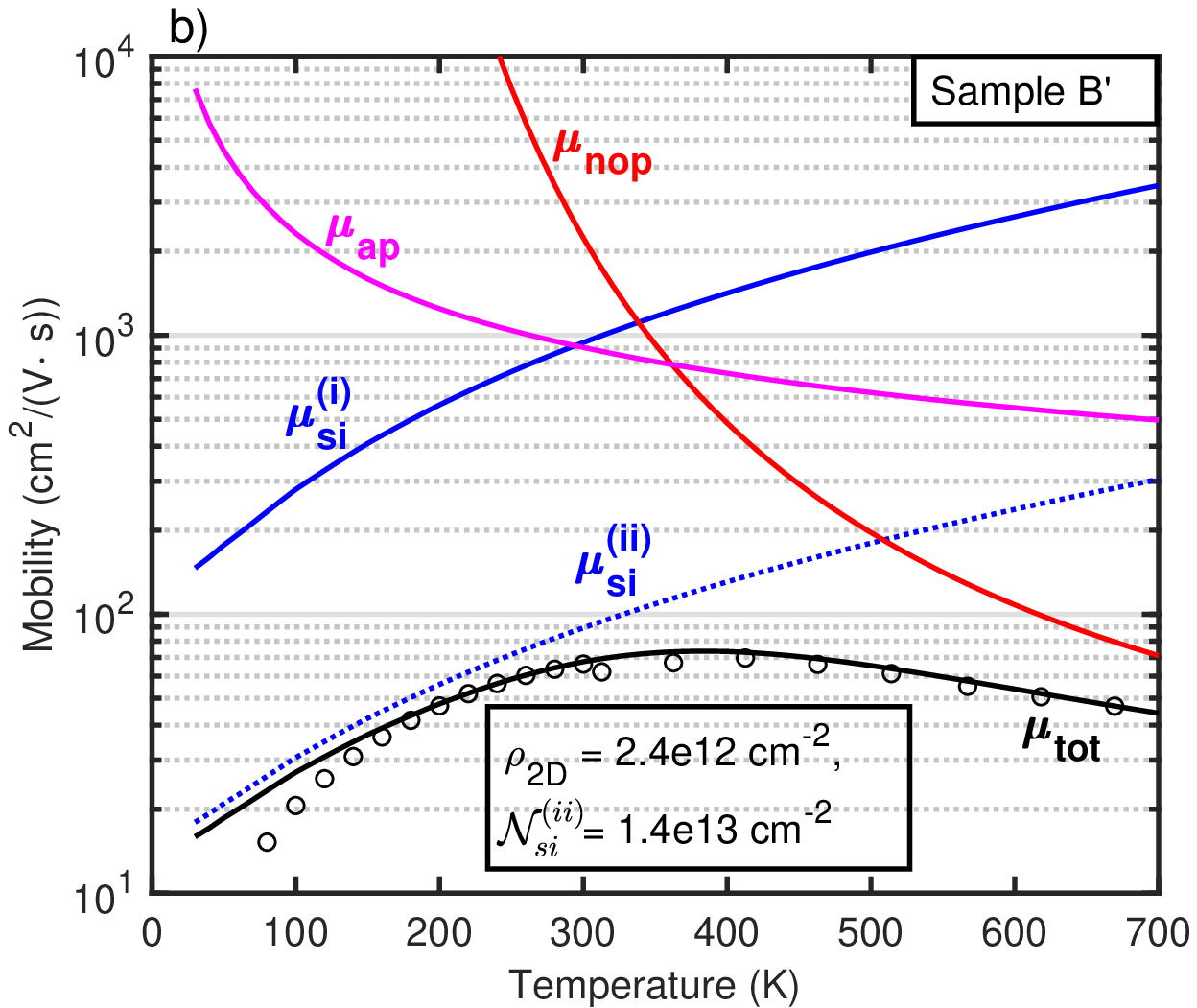}
\caption{(Color online) Measured and calculated Hall mobilities as a function of temperature for samples A' and B'. (a)~Calculations fitted to sample A' data. A moderate sheet density of 5.5$\times10^{12}$~cm$^{-2}$ yields a comparable fitting parameter for $\mathcal{N}_{si}^{(ii)}$. (b) Calculations fitted to sample B' data. A low sheet density of 2.4$\times10^{12}$~cm$^{-2}$ yields a much higher fitting parameter for $\mathcal{N}_{si}^{(ii)}$.
}
\label{mutot_SC14_SC15}
\end{figure}

A precise fitting of $\mu_{tot}$ requires treatment of individual data points. Shown in Fig.~$\ref{mutot_SC14_SC15}$ are the calculated mobilities fitted to samples A' and B'. After fabrication of the Hall-effect structures, but prior to Al$_2$O$_3$ passivation, AFM was taken on the active regions shown in Fig.~\ref{fig_xsec_sem}(b). The average measured root-mean-squared height and correlation length were $\Delta\approx 0.80\pm0.10$~nm and $\Lambda\approx60\pm10$~nm, respectively. Note that the mobility calculations for SR scattering ($\mu_{SR}$) are absent in Fig.~\ref{mutot_SC14_SC15}. This is primarily due to the $\rho_{2D}^2$ dependence of SR scattering in Eq.~(\ref{eq:M_SR2}), as well as the large $\Lambda$ measured, both of which yield an insignificant contribution to the mobility. Thus, $\mu_{SR}$ is ignored here. For higher temperatures, phonon scattering is slightly reduced for lower sheet densities, which is due to a reduction of holes occupying energy states exceeding the LO-phonon energy (and hence reducing scattering by NOP emission). As with Fig.~\ref{mutot_kasu_muSI_expPoints}(a), the coupling constant for NOP is fitted to $D_{nop} = 1.4\times10^{10}$~eV/cm. This is in close agreement with values reported for bulk diamond, where Ref. \cite{Tsukioka_2006} reported  1.2$\times10^{10}$~eV/cm and Ref. \cite{Pernot_2010} reported 0.7$\times10^{10}$~eV/cm. The AP deformation potential $D_{ap}$ was set to 8~eV, as it has also been fitted experimentally in other works for bulk diamond \cite{Pernot_2010,CARDONA1986421}.

As anticipated, SI scattering of types (i) and (ii) are dominant at low-to-intermediate temperatures (up to $\sim$450~K). This is attributed to the close proximity of the charged acceptors (i.e., $\mu_{si}^{(i)}$, $d=2.1~\mathrm{\AA}$) and the C-H disorder (i.e., $\mu_{si}^{(ii)}$, $d=0~\mathrm{\AA}$).  These calculations show that lower values of $\rho_{2D}$ require an increase in the fitting parameter $\mathcal{N}_{si}^{(ii)}$, which suggests that there is an increase in the potential fluctuations induced by C-H disorder. As discussed earlier, the nature of such fluctuations may include incomplete H-termination, a hypothesis also mentioned in Refs. \cite{Garrido_2005,Nebel_2002}. Other complex chemistry may also be involved in inducing such fluctuations, and its enhanced effect on the mobility is evident as $\rho_{2D}$ decreases. In addition to the increasing parameter $\mathcal{N}_{si}^{(ii)}$, a lower sheet density [Fig.~$\ref{mutot_SC14_SC15}$(b)] exhibits a steeper decline (and thus a higher activation energy) in both $\mu_{si}^{(i)}$ and $\mu_{si}^{(ii)}$ as T$\rightarrow$ 0~K, which is precisely what is observed in the experimental Hall data. The $\rho_{2D}$-dependent activation energy is also shown in the 1/T representation of the samples in Fig.~$\ref{rho_mu_sigE_sc14_sc19}$(b) and (c), as well as in Ref.~\cite{Garrido_2005}. This effect is attributed to the larger occupation of holes near the valence band edge ($E_V-E_F\rightarrow$~0~eV) where the scattering rate is higher.

Efforts to boost the 2D hole gas conductivity on H:Diamond surfaces must therefore attend to two design parameters. The first involves the separation of the charged surface acceptors from the 2D hole gas, evidenced by Eq. ($\ref{eq:M_SI3}$) where $\mu_{si}$ is exponentially dependent on $d$. The second is to ensure uniformity on the H:Diamond surface, which includes a complete C-H termination and a periodic distribution of the surface acceptors. However, the exact nature of the scattering associated with C-H disorder remains unclear. Thus, extensive experiments studying the surface chemistry on H:Diamond are required to find the solutions necessary to boost the conductivity and advance this promising technology.

\section{Conclusion}
We have fabricated Hall-effect structures on multiple diamond substrates with varying 2DHG conduction properties. Extensive Hall measurements were taken at temperatures ranging from 25~K to 700~K, and a scattering model was developed to explore the mobility-limiting mechanisms. A multi-band treatment of the HH, LH, and SO band was included using a Schr{\"o}dinger/Poisson solver, where only the first energy level of each band was considered. Moreover, the latest reported Luttinger parameters allowed for a parabolic treatment of the hole dispersion. The Hall measurements at low-to-intermediate temperatures suggest that long-range potential fluctuations exist, which contributes to the \textquote{ceiling} observed for the hole mobilities even at low sheet densities. These fluctuations may arise both from charged surface acceptors and disorder related to the \mbox{C-H} surface. The nature of this disorder is a subject that remains to be studied. 

\section*{Acknowledgements}
This work was supported by the National Science Foundation Graduate Research Fellowship under Grant DGE-1656518. Part of this work was performed at the Stanford Nanofabrication Facility (SNF) and the Stanford Nano Shared Facilities (SNSF). Many thanks to the SNF staff for their support. 

\nocite{*}

\bibliography{aapmsamp}

\providecommand{\noopsort}[1]{}\providecommand{\singleletter}[1]{#1}%
\begin{thebibliography}{40}%
\makeatletter
\providecommand \@ifxundefined [1]{%
 \@ifx{#1\undefined}
}%
\providecommand \@ifnum [1]{%
 \ifnum #1\expandafter \@firstoftwo
 \else \expandafter \@secondoftwo
 \fi
}%
\providecommand \@ifx [1]{%
 \ifx #1\expandafter \@firstoftwo
 \else \expandafter \@secondoftwo
 \fi
}%
\providecommand \natexlab [1]{#1}%
\providecommand \enquote  [1]{``#1''}%
\providecommand \bibnamefont  [1]{#1}%
\providecommand \bibfnamefont [1]{#1}%
\providecommand \citenamefont [1]{#1}%
\providecommand \href@noop [0]{\@secondoftwo}%
\providecommand \href [0]{\begingroup \@sanitize@url \@href}%
\providecommand \@href[1]{\@@startlink{#1}\@@href}%
\providecommand \@@href[1]{\endgroup#1\@@endlink}%
\providecommand \@sanitize@url [0]{\catcode `\\12\catcode `\$12\catcode
  `\&12\catcode `\#12\catcode `\^12\catcode `\_12\catcode `\%12\relax}%
\providecommand \@@startlink[1]{}%
\providecommand \@@endlink[0]{}%
\providecommand \url  [0]{\begingroup\@sanitize@url \@url }%
\providecommand \@url [1]{\endgroup\@href {#1}{\urlprefix }}%
\providecommand \urlprefix  [0]{URL }%
\providecommand \Eprint [0]{\href }%
\providecommand \doibase [0]{https://doi.org/}%
\providecommand \selectlanguage [0]{\@gobble}%
\providecommand \bibinfo  [0]{\@secondoftwo}%
\providecommand \bibfield  [0]{\@secondoftwo}%
\providecommand \translation [1]{[#1]}%
\providecommand \BibitemOpen [0]{}%
\providecommand \bibitemStop [0]{}%
\providecommand \bibitemNoStop [0]{.\EOS\space}%
\providecommand \EOS [0]{\spacefactor3000\relax}%
\providecommand \BibitemShut  [1]{\csname bibitem#1\endcsname}%
\let\auto@bib@innerbib\@empty
\bibitem [{\citenamefont {Geis}\ \emph {et~al.}(2018)\citenamefont {Geis},
  \citenamefont {Wade}, \citenamefont {Wuorio}, \citenamefont {Fedynyshyn},
  \citenamefont {Duncan}, \citenamefont {Plaut}, \citenamefont {Varghese},
  \citenamefont {Warnock}, \citenamefont {Vitale},\ and\ \citenamefont
  {Hollis}}]{Geis_2018}%
  \BibitemOpen
  \bibfield  {author} {\bibinfo {author} {\bibfnamefont {M.~W.}\ \bibnamefont
  {Geis}}, \bibinfo {author} {\bibfnamefont {T.~C.}\ \bibnamefont {Wade}},
  \bibinfo {author} {\bibfnamefont {C.~H.}\ \bibnamefont {Wuorio}}, \bibinfo
  {author} {\bibfnamefont {T.~H.}\ \bibnamefont {Fedynyshyn}}, \bibinfo
  {author} {\bibfnamefont {B.}~\bibnamefont {Duncan}}, \bibinfo {author}
  {\bibfnamefont {M.~E.}\ \bibnamefont {Plaut}}, \bibinfo {author}
  {\bibfnamefont {J.~O.}\ \bibnamefont {Varghese}}, \bibinfo {author}
  {\bibfnamefont {S.~M.}\ \bibnamefont {Warnock}}, \bibinfo {author}
  {\bibfnamefont {S.~A.}\ \bibnamefont {Vitale}}, and\ \bibinfo {author}
  {\bibfnamefont {M.~A.}\ \bibnamefont {Hollis}},\ }\href@noop {} {\bibfield
  {journal} {\bibinfo  {journal} {physica status solidi (a)}\ }\textbf
  {\bibinfo {volume} {215}},\ \bibinfo {pages} {1800681} (\bibinfo {year}
  {2018})}\BibitemShut {NoStop}%
\bibitem [{\citenamefont {Collins}\ and\ \citenamefont
  {Williams}(1971)}]{Collins_1971}%
  \BibitemOpen
  \bibfield  {author} {\bibinfo {author} {\bibfnamefont {A.~T.}\ \bibnamefont
  {Collins}}and\ \bibinfo {author} {\bibfnamefont {A.~W.~S.}\ \bibnamefont
  {Williams}},\ }\href@noop {} {\bibfield  {journal} {\bibinfo  {journal}
  {Journal of Physics C: Solid State Physics}\ }\textbf {\bibinfo {volume}
  {4}},\ \bibinfo {pages} {1789} (\bibinfo {year} {1971})}\BibitemShut
  {NoStop}%
\bibitem [{\citenamefont {Hirama}\ \emph {et~al.}(2010)\citenamefont {Hirama},
  \citenamefont {Tsuge}, \citenamefont {Sato}, \citenamefont {Tsuno},
  \citenamefont {Jingu}, \citenamefont {Yamauchi},\ and\ \citenamefont
  {Kawarada}}]{Hirama_2010}%
  \BibitemOpen
  \bibfield  {author} {\bibinfo {author} {\bibfnamefont {K.}~\bibnamefont
  {Hirama}}, \bibinfo {author} {\bibfnamefont {K.}~\bibnamefont {Tsuge}},
  \bibinfo {author} {\bibfnamefont {S.}~\bibnamefont {Sato}}, \bibinfo {author}
  {\bibfnamefont {T.}~\bibnamefont {Tsuno}}, \bibinfo {author} {\bibfnamefont
  {Y.}~\bibnamefont {Jingu}}, \bibinfo {author} {\bibfnamefont
  {S.}~\bibnamefont {Yamauchi}}, and\ \bibinfo {author} {\bibfnamefont
  {H.}~\bibnamefont {Kawarada}},\ }\href@noop {} {\bibfield  {journal}
  {\bibinfo  {journal} {Applied Physics Express}\ }\textbf {\bibinfo {volume}
  {3}},\ \bibinfo {pages} {044001} (\bibinfo {year} {2010})}\BibitemShut
  {NoStop}%
\bibitem [{\citenamefont {Verona}\ \emph {et~al.}(2016)\citenamefont {Verona},
  \citenamefont {Ciccognani}, \citenamefont {Colangeli}, \citenamefont
  {Limiti}, \citenamefont {Marinelli},\ and\ \citenamefont
  {Verona-Rinati}}]{verona_2016}%
  \BibitemOpen
  \bibfield  {author} {\bibinfo {author} {\bibfnamefont {C.}~\bibnamefont
  {Verona}}, \bibinfo {author} {\bibfnamefont {W.}~\bibnamefont {Ciccognani}},
  \bibinfo {author} {\bibfnamefont {S.}~\bibnamefont {Colangeli}}, \bibinfo
  {author} {\bibfnamefont {E.}~\bibnamefont {Limiti}}, \bibinfo {author}
  {\bibfnamefont {M.}~\bibnamefont {Marinelli}}, and\ \bibinfo {author}
  {\bibfnamefont {G.}~\bibnamefont {Verona-Rinati}},\ }\href@noop {} {\bibfield
   {journal} {\bibinfo  {journal} {Journal of Applied Physics}\ }\textbf
  {\bibinfo {volume} {120}},\ \bibinfo {pages} {025104} (\bibinfo {year}
  {2016})}\BibitemShut {NoStop}%
\bibitem [{\citenamefont {Maier}\ \emph {et~al.}(2000)\citenamefont {Maier},
  \citenamefont {Riedel}, \citenamefont {Mantel}, \citenamefont {Ristein},\
  and\ \citenamefont {Ley}}]{Maier_2000}%
  \BibitemOpen
  \bibfield  {author} {\bibinfo {author} {\bibfnamefont {F.}~\bibnamefont
  {Maier}}, \bibinfo {author} {\bibfnamefont {M.}~\bibnamefont {Riedel}},
  \bibinfo {author} {\bibfnamefont {B.}~\bibnamefont {Mantel}}, \bibinfo
  {author} {\bibfnamefont {J.}~\bibnamefont {Ristein}}, and\ \bibinfo {author}
  {\bibfnamefont {L.}~\bibnamefont {Ley}},\ }\href@noop {} {\bibfield
  {journal} {\bibinfo  {journal} {Phys. Rev. Lett.}\ }\textbf {\bibinfo
  {volume} {85}},\ \bibinfo {pages} {3472} (\bibinfo {year}
  {2000})}\BibitemShut {NoStop}%
\bibitem [{\citenamefont {Kawarada}(1996)}]{KAWARADA1996}%
  \BibitemOpen
  \bibfield  {author} {\bibinfo {author} {\bibfnamefont {H.}~\bibnamefont
  {Kawarada}},\ }\href@noop {} {\bibfield  {journal} {\bibinfo  {journal}
  {Surface Science Reports}\ }\textbf {\bibinfo {volume} {26}},\ \bibinfo
  {pages} {205 } (\bibinfo {year} {1996})}\BibitemShut {NoStop}%
\bibitem [{\citenamefont {Tsugawa}\ \emph {et~al.}(2001)\citenamefont
  {Tsugawa}, \citenamefont {Umezawa},\ and\ \citenamefont
  {Kawarada}}]{Tsugawa_2001}%
  \BibitemOpen
  \bibfield  {author} {\bibinfo {author} {\bibfnamefont {K.}~\bibnamefont
  {Tsugawa}}, \bibinfo {author} {\bibfnamefont {H.}~\bibnamefont {Umezawa}},
  and\ \bibinfo {author} {\bibfnamefont {H.}~\bibnamefont {Kawarada}},\
  }\href@noop {} {\bibfield  {journal} {\bibinfo  {journal} {Japanese Journal
  of Applied Physics}\ }\textbf {\bibinfo {volume} {40}},\ \bibinfo {pages}
  {3101} (\bibinfo {year} {2001})}\BibitemShut {NoStop}%
\bibitem [{\citenamefont {Riedel}\ \emph {et~al.}(2004)\citenamefont {Riedel},
  \citenamefont {Ristein},\ and\ \citenamefont {Ley}}]{Riedel_2004}%
  \BibitemOpen
  \bibfield  {author} {\bibinfo {author} {\bibfnamefont {M.}~\bibnamefont
  {Riedel}}, \bibinfo {author} {\bibfnamefont {J.}~\bibnamefont {Ristein}},
  and\ \bibinfo {author} {\bibfnamefont {L.}~\bibnamefont {Ley}},\ }\href@noop
  {} {\bibfield  {journal} {\bibinfo  {journal} {Phys. Rev. B}\ }\textbf
  {\bibinfo {volume} {69}},\ \bibinfo {pages} {125338} (\bibinfo {year}
  {2004})}\BibitemShut {NoStop}%
\bibitem [{\citenamefont {Kawarada}\ \emph {et~al.}(2014)\citenamefont
  {Kawarada}, \citenamefont {Tsuboi}, \citenamefont {Naruo}, \citenamefont
  {Yamada}, \citenamefont {Xu}, \citenamefont {Daicho}, \citenamefont {Saito},\
  and\ \citenamefont {Hiraiwa}}]{Kawarada_2014}%
  \BibitemOpen
  \bibfield  {author} {\bibinfo {author} {\bibfnamefont {H.}~\bibnamefont
  {Kawarada}}, \bibinfo {author} {\bibfnamefont {H.}~\bibnamefont {Tsuboi}},
  \bibinfo {author} {\bibfnamefont {T.}~\bibnamefont {Naruo}}, \bibinfo
  {author} {\bibfnamefont {T.}~\bibnamefont {Yamada}}, \bibinfo {author}
  {\bibfnamefont {D.}~\bibnamefont {Xu}}, \bibinfo {author} {\bibfnamefont
  {A.}~\bibnamefont {Daicho}}, \bibinfo {author} {\bibfnamefont
  {T.}~\bibnamefont {Saito}}, and\ \bibinfo {author} {\bibfnamefont
  {A.}~\bibnamefont {Hiraiwa}},\ }\href@noop {} {\bibfield  {journal} {\bibinfo
   {journal} {Applied Physics Letters}\ }\textbf {\bibinfo {volume} {105}},\
  \bibinfo {pages} {013510} (\bibinfo {year} {2014})}\BibitemShut {NoStop}%
\bibitem [{\citenamefont {Liu}\ \emph {et~al.}(2013)\citenamefont {Liu},
  \citenamefont {Liao}, \citenamefont {Imura}, \citenamefont {Oosato},
  \citenamefont {Watanabe},\ and\ \citenamefont {Koide}}]{Liu_2013}%
  \BibitemOpen
  \bibfield  {author} {\bibinfo {author} {\bibfnamefont {J.~W.}\ \bibnamefont
  {Liu}}, \bibinfo {author} {\bibfnamefont {M.~Y.}\ \bibnamefont {Liao}},
  \bibinfo {author} {\bibfnamefont {M.}~\bibnamefont {Imura}}, \bibinfo
  {author} {\bibfnamefont {H.}~\bibnamefont {Oosato}}, \bibinfo {author}
  {\bibfnamefont {E.}~\bibnamefont {Watanabe}}, and\ \bibinfo {author}
  {\bibfnamefont {Y.}~\bibnamefont {Koide}},\ }\href@noop {} {\bibfield
  {journal} {\bibinfo  {journal} {Applied Physics Letters}\ }\textbf {\bibinfo
  {volume} {102}},\ \bibinfo {pages} {112910} (\bibinfo {year}
  {2013})}\BibitemShut {NoStop}%
\bibitem [{\citenamefont {Russell}\ \emph {et~al.}(2013)\citenamefont
  {Russell}, \citenamefont {Cao}, \citenamefont {Qi}, \citenamefont {Tallaire},
  \citenamefont {Crawford}, \citenamefont {Wee},\ and\ \citenamefont
  {Moran}}]{Russel_2013}%
  \BibitemOpen
  \bibfield  {author} {\bibinfo {author} {\bibfnamefont {S.~A.~O.}\
  \bibnamefont {Russell}}, \bibinfo {author} {\bibfnamefont {L.}~\bibnamefont
  {Cao}}, \bibinfo {author} {\bibfnamefont {D.}~\bibnamefont {Qi}}, \bibinfo
  {author} {\bibfnamefont {A.}~\bibnamefont {Tallaire}}, \bibinfo {author}
  {\bibfnamefont {K.~G.}\ \bibnamefont {Crawford}}, \bibinfo {author}
  {\bibfnamefont {A.~T.~S.}\ \bibnamefont {Wee}}, and\ \bibinfo {author}
  {\bibfnamefont {D.~A.~J.}\ \bibnamefont {Moran}},\ }\href@noop {} {\bibfield
  {journal} {\bibinfo  {journal} {Applied Physics Letters}\ }\textbf {\bibinfo
  {volume} {103}},\ \bibinfo {pages} {202112} (\bibinfo {year}
  {2013})}\BibitemShut {NoStop}%
\bibitem [{\citenamefont {Rezek}\ \emph
  {et~al.}(2006{\natexlab{a}})\citenamefont {Rezek}, \citenamefont {Watanabe},\
  and\ \citenamefont {Nebel}}]{Rezek_2006}%
  \BibitemOpen
  \bibfield  {author} {\bibinfo {author} {\bibfnamefont {B.}~\bibnamefont
  {Rezek}}, \bibinfo {author} {\bibfnamefont {H.}~\bibnamefont {Watanabe}},
  and\ \bibinfo {author} {\bibfnamefont {C.~E.}\ \bibnamefont {Nebel}},\
  }\href@noop {} {\bibfield  {journal} {\bibinfo  {journal} {Applied Physics
  Letters}\ }\textbf {\bibinfo {volume} {88}},\ \bibinfo {pages} {042110}
  (\bibinfo {year} {2006}{\natexlab{a}})}\BibitemShut {NoStop}%
\bibitem [{\citenamefont {Li}\ \emph {et~al.}(2018)\citenamefont {Li},
  \citenamefont {Zhang}, \citenamefont {Liu}, \citenamefont {Ren},
  \citenamefont {Zhang},\ and\ \citenamefont {Hao}}]{YLIetal}%
  \BibitemOpen
  \bibfield  {author} {\bibinfo {author} {\bibfnamefont {Y.}~\bibnamefont
  {Li}}, \bibinfo {author} {\bibfnamefont {J.-F.}\ \bibnamefont {Zhang}},
  \bibinfo {author} {\bibfnamefont {G.-P.}\ \bibnamefont {Liu}}, \bibinfo
  {author} {\bibfnamefont {Z.-Y.}\ \bibnamefont {Ren}}, \bibinfo {author}
  {\bibfnamefont {J.-C.}\ \bibnamefont {Zhang}}, and\ \bibinfo {author}
  {\bibfnamefont {Y.}~\bibnamefont {Hao}},\ }\href@noop {} {\bibfield
  {journal} {\bibinfo  {journal} {physica status solidi (RRL)}\ }\textbf
  {\bibinfo {volume} {12}},\ \bibinfo {pages} {1700401} (\bibinfo {year}
  {2018})}\BibitemShut {NoStop}%
\bibitem [{\citenamefont {Nebel}\ \emph {et~al.}(2002)\citenamefont {Nebel},
  \citenamefont {Ertl}, \citenamefont {Sauerer}, \citenamefont {Stutzmann},
  \citenamefont {Graeff}, \citenamefont {Bergonzo}, \citenamefont {Williams},\
  and\ \citenamefont {Jackman}}]{Nebel_2002}%
  \BibitemOpen
  \bibfield  {author} {\bibinfo {author} {\bibfnamefont {C.}~\bibnamefont
  {Nebel}}, \bibinfo {author} {\bibfnamefont {F.}~\bibnamefont {Ertl}},
  \bibinfo {author} {\bibfnamefont {C.}~\bibnamefont {Sauerer}}, \bibinfo
  {author} {\bibfnamefont {M.}~\bibnamefont {Stutzmann}}, \bibinfo {author}
  {\bibfnamefont {C.}~\bibnamefont {Graeff}}, \bibinfo {author} {\bibfnamefont
  {P.}~\bibnamefont {Bergonzo}}, \bibinfo {author} {\bibfnamefont
  {O.}~\bibnamefont {Williams}}, and\ \bibinfo {author} {\bibfnamefont
  {R.}~\bibnamefont {Jackman}},\ }\href@noop {} {\bibfield  {journal} {\bibinfo
   {journal} {Diamond and Related Materials}\ }\textbf {\bibinfo {volume}
  {11}},\ \bibinfo {pages} {351 } (\bibinfo {year} {2002})}\BibitemShut
  {NoStop}%
\bibitem [{\citenamefont {Garrido}\ \emph {et~al.}(2005)\citenamefont
  {Garrido}, \citenamefont {Heimbeck},\ and\ \citenamefont
  {Stutzmann}}]{Garrido_2005}%
  \BibitemOpen
  \bibfield  {author} {\bibinfo {author} {\bibfnamefont {J.~A.}\ \bibnamefont
  {Garrido}}, \bibinfo {author} {\bibfnamefont {T.}~\bibnamefont {Heimbeck}},
  and\ \bibinfo {author} {\bibfnamefont {M.}~\bibnamefont {Stutzmann}},\
  }\href@noop {} {\bibfield  {journal} {\bibinfo  {journal} {Phys. Rev. B}\
  }\textbf {\bibinfo {volume} {71}},\ \bibinfo {pages} {245310} (\bibinfo
  {year} {2005})}\BibitemShut {NoStop}%
\bibitem [{\citenamefont {Kasu}\ \emph {et~al.}(2012)\citenamefont {Kasu},
  \citenamefont {Sato},\ and\ \citenamefont {Hirama}}]{Kasu_2012}%
  \BibitemOpen
  \bibfield  {author} {\bibinfo {author} {\bibfnamefont {M.}~\bibnamefont
  {Kasu}}, \bibinfo {author} {\bibfnamefont {H.}~\bibnamefont {Sato}}, and\
  \bibinfo {author} {\bibfnamefont {K.}~\bibnamefont {Hirama}},\ }\href@noop {}
  {\bibfield  {journal} {\bibinfo  {journal} {Applied Physics Express}\
  }\textbf {\bibinfo {volume} {5}},\ \bibinfo {pages} {025701} (\bibinfo {year}
  {2012})}\BibitemShut {NoStop}%
\bibitem [{\citenamefont {{Birner}}\ \emph {et~al.}(2007)\citenamefont
  {{Birner}}, \citenamefont {{Zibold}}, \citenamefont {{Andlauer}},
  \citenamefont {{Kubis}}, \citenamefont {{Sabathil}}, \citenamefont
  {{Trellakis}},\ and\ \citenamefont {{Vogl}}}]{nextnano1}%
  \BibitemOpen
  \bibfield  {author} {\bibinfo {author} {\bibfnamefont {S.}~\bibnamefont
  {{Birner}}}, \bibinfo {author} {\bibfnamefont {T.}~\bibnamefont {{Zibold}}},
  \bibinfo {author} {\bibfnamefont {T.}~\bibnamefont {{Andlauer}}}, \bibinfo
  {author} {\bibfnamefont {T.}~\bibnamefont {{Kubis}}}, \bibinfo {author}
  {\bibfnamefont {M.}~\bibnamefont {{Sabathil}}}, \bibinfo {author}
  {\bibfnamefont {A.}~\bibnamefont {{Trellakis}}}, and\ \bibinfo {author}
  {\bibfnamefont {P.}~\bibnamefont {{Vogl}}},\ }\href@noop {} {\bibfield
  {journal} {\bibinfo  {journal} {IEEE Transactions on Electron Devices}\
  }\textbf {\bibinfo {volume} {54}},\ \bibinfo {pages} {2137} (\bibinfo {year}
  {2007})}\BibitemShut {NoStop}%
\bibitem [{\citenamefont {Dankerl}\ \emph {et~al.}(2011)\citenamefont
  {Dankerl}, \citenamefont {Lippert}, \citenamefont {Birner}, \citenamefont
  {St\"utzel}, \citenamefont {Stutzmann},\ and\ \citenamefont
  {Garrido}}]{dia_nn1}%
  \BibitemOpen
  \bibfield  {author} {\bibinfo {author} {\bibfnamefont {M.}~\bibnamefont
  {Dankerl}}, \bibinfo {author} {\bibfnamefont {A.}~\bibnamefont {Lippert}},
  \bibinfo {author} {\bibfnamefont {S.}~\bibnamefont {Birner}}, \bibinfo
  {author} {\bibfnamefont {E.~U.}\ \bibnamefont {St\"utzel}}, \bibinfo {author}
  {\bibfnamefont {M.}~\bibnamefont {Stutzmann}}, and\ \bibinfo {author}
  {\bibfnamefont {J.~A.}\ \bibnamefont {Garrido}},\ }\href@noop {} {\bibfield
  {journal} {\bibinfo  {journal} {Phys. Rev. Lett.}\ }\textbf {\bibinfo
  {volume} {106}},\ \bibinfo {pages} {196103} (\bibinfo {year}
  {2011})}\BibitemShut {NoStop}%
\bibitem [{\citenamefont {Newell}\ \emph {et~al.}(2016)\citenamefont {Newell},
  \citenamefont {Dowdell},\ and\ \citenamefont {Santamore}}]{dia_nn2}%
  \BibitemOpen
  \bibfield  {author} {\bibinfo {author} {\bibfnamefont {A.~N.}\ \bibnamefont
  {Newell}}, \bibinfo {author} {\bibfnamefont {D.~A.}\ \bibnamefont {Dowdell}},
  and\ \bibinfo {author} {\bibfnamefont {D.~H.}\ \bibnamefont {Santamore}},\
  }\href@noop {} {\bibfield  {journal} {\bibinfo  {journal} {Journal of Applied
  Physics}\ }\textbf {\bibinfo {volume} {120}} (\bibinfo {year}
  {2016})}\BibitemShut {NoStop}%
\bibitem [{\citenamefont {Naka}\ \emph {et~al.}(2013)\citenamefont {Naka},
  \citenamefont {Fukai}, \citenamefont {Handa},\ and\ \citenamefont
  {Akimoto}}]{NAKA}%
  \BibitemOpen
  \bibfield  {author} {\bibinfo {author} {\bibfnamefont {N.}~\bibnamefont
  {Naka}}, \bibinfo {author} {\bibfnamefont {K.}~\bibnamefont {Fukai}},
  \bibinfo {author} {\bibfnamefont {Y.}~\bibnamefont {Handa}}, and\ \bibinfo
  {author} {\bibfnamefont {I.}~\bibnamefont {Akimoto}},\ }\href@noop {}
  {\bibfield  {journal} {\bibinfo  {journal} {Phys. Rev. B}\ }\textbf {\bibinfo
  {volume} {88}},\ \bibinfo {pages} {035205} (\bibinfo {year}
  {2013})}\BibitemShut {NoStop}%
\bibitem [{\citenamefont {Rauch}(1961)}]{SO_BAND_exp}%
  \BibitemOpen
  \bibfield  {author} {\bibinfo {author} {\bibfnamefont {C.~J.}\ \bibnamefont
  {Rauch}},\ }\href@noop {} {\bibfield  {journal} {\bibinfo  {journal} {Phys.
  Rev. Lett.}\ }\textbf {\bibinfo {volume} {7}},\ \bibinfo {pages} {83}
  (\bibinfo {year} {1961})}\BibitemShut {NoStop}%
\bibitem [{\citenamefont {Dresselhaus}\ \emph {et~al.}(1955)\citenamefont
  {Dresselhaus}, \citenamefont {Kip},\ and\ \citenamefont {Kittel}}]{EK_DISP}%
  \BibitemOpen
  \bibfield  {author} {\bibinfo {author} {\bibfnamefont {G.}~\bibnamefont
  {Dresselhaus}}, \bibinfo {author} {\bibfnamefont {A.~F.}\ \bibnamefont
  {Kip}}, and\ \bibinfo {author} {\bibfnamefont {C.}~\bibnamefont {Kittel}},\
  }\href@noop {} {\bibfield  {journal} {\bibinfo  {journal} {Phys. Rev.}\
  }\textbf {\bibinfo {volume} {98}},\ \bibinfo {pages} {368} (\bibinfo {year}
  {1955})}\BibitemShut {NoStop}%
\bibitem [{\citenamefont {Kono}\ \emph {et~al.}(1993)\citenamefont {Kono},
  \citenamefont {Takeyama}, \citenamefont {Takamasu}, \citenamefont {Miura},
  \citenamefont {Fujimori}, \citenamefont {Nishibayashi}, \citenamefont
  {Nakajima},\ and\ \citenamefont {Tsuji}}]{meff_old1}%
  \BibitemOpen
  \bibfield  {author} {\bibinfo {author} {\bibfnamefont {J.}~\bibnamefont
  {Kono}}, \bibinfo {author} {\bibfnamefont {S.}~\bibnamefont {Takeyama}},
  \bibinfo {author} {\bibfnamefont {T.}~\bibnamefont {Takamasu}}, \bibinfo
  {author} {\bibfnamefont {N.}~\bibnamefont {Miura}}, \bibinfo {author}
  {\bibfnamefont {N.}~\bibnamefont {Fujimori}}, \bibinfo {author}
  {\bibfnamefont {Y.}~\bibnamefont {Nishibayashi}}, \bibinfo {author}
  {\bibfnamefont {T.}~\bibnamefont {Nakajima}}, and\ \bibinfo {author}
  {\bibfnamefont {K.}~\bibnamefont {Tsuji}},\ }\href@noop {} {\bibfield
  {journal} {\bibinfo  {journal} {Phys. Rev. B}\ }\textbf {\bibinfo {volume}
  {48}},\ \bibinfo {pages} {10917} (\bibinfo {year} {1993})}\BibitemShut
  {NoStop}%
\bibitem [{\citenamefont {Willatzen}\ \emph {et~al.}(1994)\citenamefont
  {Willatzen}, \citenamefont {Cardona},\ and\ \citenamefont
  {Christensen}}]{meff_old2}%
  \BibitemOpen
  \bibfield  {author} {\bibinfo {author} {\bibfnamefont {M.}~\bibnamefont
  {Willatzen}}, \bibinfo {author} {\bibfnamefont {M.}~\bibnamefont {Cardona}},
  and\ \bibinfo {author} {\bibfnamefont {N.~E.}\ \bibnamefont {Christensen}},\
  }\href@noop {} {\bibfield  {journal} {\bibinfo  {journal} {Phys. Rev. B}\
  }\textbf {\bibinfo {volume} {50}},\ \bibinfo {pages} {18054} (\bibinfo {year}
  {1994})}\BibitemShut {NoStop}%
\bibitem [{\citenamefont {Takahide}\ \emph {et~al.}(2014)\citenamefont
  {Takahide}, \citenamefont {Okazaki}, \citenamefont {Deguchi}, \citenamefont
  {Uji}, \citenamefont {Takeya}, \citenamefont {Takano}, \citenamefont
  {Tsuboi},\ and\ \citenamefont {Kawarada}}]{Takahide}%
  \BibitemOpen
  \bibfield  {author} {\bibinfo {author} {\bibfnamefont {Y.}~\bibnamefont
  {Takahide}}, \bibinfo {author} {\bibfnamefont {H.}~\bibnamefont {Okazaki}},
  \bibinfo {author} {\bibfnamefont {K.}~\bibnamefont {Deguchi}}, \bibinfo
  {author} {\bibfnamefont {S.}~\bibnamefont {Uji}}, \bibinfo {author}
  {\bibfnamefont {H.}~\bibnamefont {Takeya}}, \bibinfo {author} {\bibfnamefont
  {Y.}~\bibnamefont {Takano}}, \bibinfo {author} {\bibfnamefont
  {H.}~\bibnamefont {Tsuboi}}, and\ \bibinfo {author} {\bibfnamefont
  {H.}~\bibnamefont {Kawarada}},\ }\href@noop {} {\bibfield  {journal}
  {\bibinfo  {journal} {Phys. Rev. B}\ }\textbf {\bibinfo {volume} {89}},\
  \bibinfo {pages} {235304} (\bibinfo {year} {2014})}\BibitemShut {NoStop}%
\bibitem [{\citenamefont {Davies}(1997)}]{davies_1997}%
  \BibitemOpen
  \bibfield  {author} {\bibinfo {author} {\bibfnamefont {J.~H.}\ \bibnamefont
  {Davies}},\ }in\ \href@noop {} {\emph {\bibinfo {booktitle} {The Physics of
  Low-dimensional Semiconductors: An Introduction}}}\ (\bibinfo  {publisher}
  {Cambridge University Press},\ \bibinfo {year} {1997})\ pp.\ \bibinfo {pages}
  {329--370}\BibitemShut {NoStop}%
\bibitem [{\citenamefont {Sato}\ and\ \citenamefont {Kasu}(2012)}]{SATO2012}%
  \BibitemOpen
  \bibfield  {author} {\bibinfo {author} {\bibfnamefont {H.}~\bibnamefont
  {Sato}}and\ \bibinfo {author} {\bibfnamefont {M.}~\bibnamefont {Kasu}},\
  }\href@noop {} {\bibfield  {journal} {\bibinfo  {journal} {Diamond and
  Related Materials}\ }\textbf {\bibinfo {volume} {24}},\ \bibinfo {pages} {99}
  (\bibinfo {year} {2012})}\BibitemShut {NoStop}%
\bibitem [{\citenamefont {Ando}\ \emph {et~al.}(1982)\citenamefont {Ando},
  \citenamefont {Fowler},\ and\ \citenamefont {Stern}}]{ando_1982}%
  \BibitemOpen
  \bibfield  {author} {\bibinfo {author} {\bibfnamefont {T.}~\bibnamefont
  {Ando}}, \bibinfo {author} {\bibfnamefont {A.~B.}\ \bibnamefont {Fowler}},
  and\ \bibinfo {author} {\bibfnamefont {F.}~\bibnamefont {Stern}},\
  }\href@noop {} {\bibfield  {journal} {\bibinfo  {journal} {Rev. Mod. Phys.}\
  }\textbf {\bibinfo {volume} {54}},\ \bibinfo {pages} {437} (\bibinfo {year}
  {1982})}\BibitemShut {NoStop}%
\bibitem [{\citenamefont {Hamaguchi}(2017)}]{Hamaguchi2017}%
  \BibitemOpen
  \bibfield  {author} {\bibinfo {author} {\bibfnamefont {C.}~\bibnamefont
  {Hamaguchi}},\ }in\ \href@noop {} {\emph {\bibinfo {booktitle} {Basic
  Semiconductor Physics}}}\ (\bibinfo  {publisher} {Springer International
  Publishing},\ \bibinfo {year} {2017})\ pp.\ \bibinfo {pages}
  {265--364}\BibitemShut {NoStop}%
\bibitem [{\citenamefont {Hirama}\ \emph {et~al.}(2008)\citenamefont {Hirama},
  \citenamefont {Takayanagi}, \citenamefont {Yamauchi}, \citenamefont {Yang},
  \citenamefont {Kawarada},\ and\ \citenamefont {Umezawa}}]{Hirama_2008}%
  \BibitemOpen
  \bibfield  {author} {\bibinfo {author} {\bibfnamefont {K.}~\bibnamefont
  {Hirama}}, \bibinfo {author} {\bibfnamefont {H.}~\bibnamefont {Takayanagi}},
  \bibinfo {author} {\bibfnamefont {S.}~\bibnamefont {Yamauchi}}, \bibinfo
  {author} {\bibfnamefont {J.~H.}\ \bibnamefont {Yang}}, \bibinfo {author}
  {\bibfnamefont {H.}~\bibnamefont {Kawarada}}, and\ \bibinfo {author}
  {\bibfnamefont {H.}~\bibnamefont {Umezawa}},\ }\href@noop {} {\bibfield
  {journal} {\bibinfo  {journal} {Applied Physics Letters}\ }\textbf {\bibinfo
  {volume} {92}},\ \bibinfo {pages} {112107} (\bibinfo {year}
  {2008})}\BibitemShut {NoStop}%
\bibitem [{\citenamefont {Crawford}\ \emph {et~al.}(2018)\citenamefont
  {Crawford}, \citenamefont {Tallaire}, \citenamefont {Li}, \citenamefont
  {Macdonald}, \citenamefont {Qi},\ and\ \citenamefont {Moran}}]{CRAWFORD2018}%
  \BibitemOpen
  \bibfield  {author} {\bibinfo {author} {\bibfnamefont {K.~G.}\ \bibnamefont
  {Crawford}}, \bibinfo {author} {\bibfnamefont {A.}~\bibnamefont {Tallaire}},
  \bibinfo {author} {\bibfnamefont {X.}~\bibnamefont {Li}}, \bibinfo {author}
  {\bibfnamefont {D.~A.}\ \bibnamefont {Macdonald}}, \bibinfo {author}
  {\bibfnamefont {D.}~\bibnamefont {Qi}}, and\ \bibinfo {author} {\bibfnamefont
  {D.~A.}\ \bibnamefont {Moran}},\ }\href@noop {} {\bibfield  {journal}
  {\bibinfo  {journal} {Diamond and Related Materials}\ }\textbf {\bibinfo
  {volume} {84}},\ \bibinfo {pages} {48} (\bibinfo {year} {2018})}\BibitemShut
  {NoStop}%
\bibitem [{\citenamefont {Oing}\ \emph {et~al.}(2019)\citenamefont {Oing},
  \citenamefont {Geller}, \citenamefont {Lorke},\ and\ \citenamefont
  {WÃ¶hrl}}]{OING2019}%
  \BibitemOpen
  \bibfield  {author} {\bibinfo {author} {\bibfnamefont {D.}~\bibnamefont
  {Oing}}, \bibinfo {author} {\bibfnamefont {M.}~\bibnamefont {Geller}},
  \bibinfo {author} {\bibfnamefont {A.}~\bibnamefont {Lorke}}, and\ \bibinfo
  {author} {\bibfnamefont {N.}~\bibnamefont {WÃ¶hrl}},\ }\href@noop {}
  {\bibfield  {journal} {\bibinfo  {journal} {Diamond and Related Materials}\
  }\textbf {\bibinfo {volume} {97}},\ \bibinfo {pages} {107450} (\bibinfo
  {year} {2019})}\BibitemShut {NoStop}%
\bibitem [{\citenamefont {Liu}\ \emph {et~al.}(2015)\citenamefont {Liu},
  \citenamefont {Cui}, \citenamefont {Qu},\ and\ \citenamefont
  {Di}}]{Liu_Fengbin_2015}%
  \BibitemOpen
  \bibfield  {author} {\bibinfo {author} {\bibfnamefont {F.}~\bibnamefont
  {Liu}}, \bibinfo {author} {\bibfnamefont {Y.}~\bibnamefont {Cui}}, \bibinfo
  {author} {\bibfnamefont {M.}~\bibnamefont {Qu}}, and\ \bibinfo {author}
  {\bibfnamefont {J.}~\bibnamefont {Di}},\ }\href@noop {} {\bibfield  {journal}
  {\bibinfo  {journal} {AIP Advances}\ }\textbf {\bibinfo {volume} {5}},\
  \bibinfo {pages} {041307} (\bibinfo {year} {2015})}\BibitemShut {NoStop}%
\bibitem [{\citenamefont {Pernot}\ \emph {et~al.}(2010)\citenamefont {Pernot},
  \citenamefont {Volpe}, \citenamefont {Omn\`es}, \citenamefont {Muret},
  \citenamefont {Mortet}, \citenamefont {Haenen},\ and\ \citenamefont
  {Teraji}}]{Pernot_2010}%
  \BibitemOpen
  \bibfield  {author} {\bibinfo {author} {\bibfnamefont {J.}~\bibnamefont
  {Pernot}}, \bibinfo {author} {\bibfnamefont {P.~N.}\ \bibnamefont {Volpe}},
  \bibinfo {author} {\bibfnamefont {F.}~\bibnamefont {Omn\`es}}, \bibinfo
  {author} {\bibfnamefont {P.}~\bibnamefont {Muret}}, \bibinfo {author}
  {\bibfnamefont {V.}~\bibnamefont {Mortet}}, \bibinfo {author} {\bibfnamefont
  {K.}~\bibnamefont {Haenen}}, and\ \bibinfo {author} {\bibfnamefont
  {T.}~\bibnamefont {Teraji}},\ }\href@noop {} {\bibfield  {journal} {\bibinfo
  {journal} {Phys. Rev. B}\ }\textbf {\bibinfo {volume} {81}},\ \bibinfo
  {pages} {205203} (\bibinfo {year} {2010})}\BibitemShut {NoStop}%
\bibitem [{\citenamefont {Ren}\ \emph {et~al.}(2018)\citenamefont {Ren},
  \citenamefont {Yuan}, \citenamefont {Zhang}, \citenamefont {Xu},
  \citenamefont {Zhang}, \citenamefont {Chen},\ and\ \citenamefont
  {Hao}}]{Ren_2018}%
  \BibitemOpen
  \bibfield  {author} {\bibinfo {author} {\bibfnamefont {Z.}~\bibnamefont
  {Ren}}, \bibinfo {author} {\bibfnamefont {G.}~\bibnamefont {Yuan}}, \bibinfo
  {author} {\bibfnamefont {J.}~\bibnamefont {Zhang}}, \bibinfo {author}
  {\bibfnamefont {L.}~\bibnamefont {Xu}}, \bibinfo {author} {\bibfnamefont
  {J.}~\bibnamefont {Zhang}}, \bibinfo {author} {\bibfnamefont
  {W.}~\bibnamefont {Chen}}, and\ \bibinfo {author} {\bibfnamefont
  {Y.}~\bibnamefont {Hao}},\ }\href@noop {} {\bibfield  {journal} {\bibinfo
  {journal} {AIP Advances}\ }\textbf {\bibinfo {volume} {8}},\ \bibinfo {pages}
  {065026} (\bibinfo {year} {2018})}\BibitemShut {NoStop}%
\bibitem [{\citenamefont {Arnold}(1976)}]{ARNOLD_1976}%
  \BibitemOpen
  \bibfield  {author} {\bibinfo {author} {\bibfnamefont {E.}~\bibnamefont
  {Arnold}},\ }\href@noop {} {\bibfield  {journal} {\bibinfo  {journal}
  {Surface Science}\ }\textbf {\bibinfo {volume} {58}},\ \bibinfo {pages} {60 }
  (\bibinfo {year} {1976})}\BibitemShut {NoStop}%
\bibitem [{\citenamefont {Ando}\ \emph {et~al.}(1993)\citenamefont {Ando},
  \citenamefont {Yamamoto}, \citenamefont {Ishii}, \citenamefont {Kamo},\ and\
  \citenamefont {Sato}}]{Ando_1993}%
  \BibitemOpen
  \bibfield  {author} {\bibinfo {author} {\bibfnamefont {T.}~\bibnamefont
  {Ando}}, \bibinfo {author} {\bibfnamefont {K.}~\bibnamefont {Yamamoto}},
  \bibinfo {author} {\bibfnamefont {M.}~\bibnamefont {Ishii}}, \bibinfo
  {author} {\bibfnamefont {M.}~\bibnamefont {Kamo}}, and\ \bibinfo {author}
  {\bibfnamefont {Y.}~\bibnamefont {Sato}},\ }\href@noop {} {\bibfield
  {journal} {\bibinfo  {journal} {J. Chem. Soc.{,} Faraday Trans.}\ }\textbf
  {\bibinfo {volume} {89}},\ \bibinfo {pages} {3635} (\bibinfo {year}
  {1993})}\BibitemShut {NoStop}%
\bibitem [{\citenamefont {Tsukioka}\ and\ \citenamefont
  {Okushi}(2006)}]{Tsukioka_2006}%
  \BibitemOpen
  \bibfield  {author} {\bibinfo {author} {\bibfnamefont {K.}~\bibnamefont
  {Tsukioka}}and\ \bibinfo {author} {\bibfnamefont {H.}~\bibnamefont
  {Okushi}},\ }\href@noop {} {\bibfield  {journal} {\bibinfo  {journal}
  {Japanese Journal of Applied Physics}\ }\textbf {\bibinfo {volume} {45}},\
  \bibinfo {pages} {8571} (\bibinfo {year} {2006})}\BibitemShut {NoStop}%
\bibitem [{\citenamefont {Cardona}\ and\ \citenamefont
  {Christensen}(1986)}]{CARDONA1986421}%
  \BibitemOpen
  \bibfield  {author} {\bibinfo {author} {\bibfnamefont {M.}~\bibnamefont
  {Cardona}}and\ \bibinfo {author} {\bibfnamefont {N.}~\bibnamefont
  {Christensen}},\ }\href@noop {} {\bibfield  {journal} {\bibinfo  {journal}
  {Solid State Communications}\ }\textbf {\bibinfo {volume} {58}},\ \bibinfo
  {pages} {421 } (\bibinfo {year} {1986})}\BibitemShut {NoStop}%
\bibitem [{\citenamefont {Rezek}\ \emph
  {et~al.}(2006{\natexlab{b}})\citenamefont {Rezek}, \citenamefont {Watanabe},\
  and\ \citenamefont {Nebel}}]{Nebel_2006}%
  \BibitemOpen
  \bibfield  {author} {\bibinfo {author} {\bibfnamefont {B.}~\bibnamefont
  {Rezek}}, \bibinfo {author} {\bibfnamefont {H.}~\bibnamefont {Watanabe}},
  and\ \bibinfo {author} {\bibfnamefont {C.~E.}\ \bibnamefont {Nebel}},\
  }\href@noop {} {\bibfield  {journal} {\bibinfo  {journal} {Applied Physics
  Letters}\ }\textbf {\bibinfo {volume} {88}},\ \bibinfo {pages} {042110}
  (\bibinfo {year} {2006}{\natexlab{b}})}\BibitemShut {NoStop}%
\end{thebibliography}%

\end{document}